\documentclass[acmsmall, screen, nonacm]{acmart}
\usepackage{longtable}
\usepackage{rotating}
\usepackage{multirow}
\usepackage{threeparttable}
\usepackage{csquotes}
\usepackage{siunitx}
\usepackage{tikz}
\usetikzlibrary{arrows.meta, positioning, calc}
\acmJournal{JDIQ}
\copyrightyear{2026}
\acmYear{2026}
\begin{document}

\def\sectionautorefname{Section}
\def\subsectionautorefname{Section}
\def\subsubsectionautorefname{Section}

\title{How to Define the Quality of Data? A Feature-Based Literature Survey}

\author{Markus Matoni}
\orcid{0000-0003-4389-5871}
\affiliation{
    \institution{Gesellschaft für wissenschaftliche Datenverarbeitung mbH Göttingen}
    \city{Göttingen}
    \country{Germany}
}
\email{markus.matoni@gwdg.de}

\author{Arno Kesper}
\orcid{0000-0002-5042-1087}
\affiliation{
    \institution{Philipps-Universität Marburg}
    \city{Marburg}
    \country{Germany}
}
\email{arno.kesper@uni-marburg.de}

\author{Gabriele Taentzer}
\orcid{0000-0002-3975-5238}
\affiliation{
    \institution{Philipps-Universität Marburg}
    \city{Marburg}
    \country{Germany}
}
\email{taentzer@uni-marburg.de}

\begin{abstract}
    %!TEX root = ../main.tex
Data is generated in almost every digital interaction, and using it effectively requires high quality.
But what exactly is data quality? The literature offers many, widely diverging answers.
This article presents a systematic literature review that screened more than 8,700 records on data quality and identified 35 providing original, dimension-based definitions.
Using feature-oriented domain analysis, a feature model of these heterogeneous definitions is derived and used to classify all identified publications.
The classification reveals a heterogeneous landscape: despite an ISO standard and a few highly cited quasi-standards, no consensus exists on which quality dimensions to consider or how to define them.
As definitions increasingly emerge for specific domains, contexts of use, and new types of data, the respective communities need to deliberately converge on how data quality is defined in their field.

\end{abstract}

%% Generated by the tool at http://dl.acm.org/ccs.cfm



% \received{21 December 2024}
% \received[revised]{}
% \received[accepted]{}

\maketitle

\section{Introduction}
%!TEX root = ../main.tex
\label{sec:introduction}
In today's data-driven world, the ability to derive insights and make informed decisions relies heavily on the quality of available data.
As data continues to proliferate across domains, ensuring its quality has become a fundamental challenge.
However, the quality of data can vary considerably.
For example, data can contain inconsistencies or incomplete information, also known as quality problems.
In order to enable high-quality data analysis and processing, a key challenge is to define the quality of the data and make this information explicit.
But how exactly is data quality defined?

The definition of data quality (DQ) has been a significant research topic for decades, which has led to a number of DQ definitions.
Existing definitions vary widely in their scope and focus.
Some articles, such as Wang \& Strong~\cite{wang1996} and ISO 25012~\cite{iso2008}, define DQ as a multifaceted concept consisting of a set of quality dimensions, with each dimension describing a particular aspect of DQ.
While these articles define DQ for data in general,
others are specific to certain formats, such as linked data (e.g., Zaveri et al.~\cite{zaveri2015}), or to certain domains, such as healthcare (e.g., Chiasera et al.~\cite{chiasera2011}).

Collectively, a set of quality dimensions, such as accuracy, completeness, and timeliness, serve as a conceptual framework for DQ. However, the existing DQ definitions vary widely in terms of both the dimensions they capture and the way they define these dimensions.
Quality dimensions are often clustered into groups, such as intrinsic, contextual, and representational DQ (cf. Wang \& Strong~\cite{wang1996}), as a try to classify different definitions of DQ.
\emph{Despite some similarities in the definitions presented in the literature, there is no consensus on a single, generally accepted definition of DQ.}

A considerable number of literature surveys have already been published on the definition of DQ.
These include surveys by Scannapieco et al.~\cite{scannapieco2002}, Sidi et al.~\cite{sidi2012}, Laranjeiro et al.~\cite{laranjeiro2015}, Zaveri et al.~\cite{zaveri2015}, Cichy et al.~\cite{cichy2019}, Wang et al.~\cite{wang2024}, and Mohammed et al.~\cite{mohammed2024}.
However, only a few surveys, such as the one by Zaveri et al.~\cite{zaveri2015}, are systematic in the sense that they present objective selection criteria for the publications to be reviewed.
In this case, the definition is specific to the quality of linked data and not to the quality of data in general.
Another systematic survey was conducted by Wang et al.~\cite{wang1995framework}.
However, it was a very early work that included only publications prior to 1995.
\emph{Consequently, there is a lack of a systematic and up-to-date literature review on the definition of DQ in general.}

Existing literature surveys on DQ definitions typically compare and categorize quality dimensions.
However, the definitions for each individual dimension can vary considerably from one publication to another, making it impossible to categorize DQ definitions by quality dimensions.
Furthermore, \emph{none of the surveys used a classification framework that allows researchers to appropriately position very different DQ definitions.}\\

In this article, we present a meta-study in which we screened more than \num{8700} records on DQ and identified 35 publications as defining DQ.
We classified these 35 publications to get a \emph{detailed understanding of how DQ is defined in the literature}, both in general and in more specific contexts.
Based on this work, we aim \emph{to identify research gaps in defining DQ}. To this end, the following research questions are addressed in this paper:
\begin{enumerate}
    \item[(RQ1)] Which publications contain original, dimension-based definitions of data quality?
    \item[(RQ2)] How can the data quality definitions found in the literature be classified?
    \item[(RQ3)] What research gaps can be identified?
\end{enumerate}

To answer these questions, this work makes the following key contributions to the field of DQ research:
\begin{enumerate}
    \item
    \emph{Systematic Literature Review on Data Quality Definitions}: We conduct a comprehensive systematic literature review to examine existing definitions of DQ based on quality dimensions.
    Our analysis includes definitions tailored to specific data representations and domains as well as those applicable to general, non-specific data contexts.
    \item
    \emph{Feature-Oriented Domain Analysis for Data Quality}: To categorize and analyze the literature, we use Feature-Oriented Domain Analysis (FODA) \cite{kang1990}.
    FODA identifies the commonalities and differences of concepts in a given domain and represents them as a feature model.
    We use FODA to specify a structured feature model of DQ definitions.
    Using our proposed feature model, we systematically classify all previously found publications on DQ definition.
    \item
    \emph{Research Gap Analysis}:
    The classification of publications provides a clear organizational structure for existing DQ definitions and helps to identify critical research gaps with respect to DQ definition.
\end{enumerate}

The contributions of this paper serve as a guide for researchers and practitioners in advancing the definition and assessment of DQ by capturing the state of the art in the literature (RQ 1). Furthermore, we contribute a feature model of DQ definitions and use it to classify the literature found (RQ 2). By comparing the existing publications on the definition of DQ (RQ 3), we provide an overview of how DQ is defined and what researchers and practitioners can build on to advance the definition of DQ in the future.

This paper is organized as follows: We summarize the related work, i.e., existing literature reviews in the context of DQ, in \autoref{sec:related_work}. In \autoref{sec:slr}, we present the results of our systematic literature review to capture the state of the art in defining DQ and answer (RQ 1). In \autoref{sec:foda}, we present our feature model for DQ definitions and classify the literature found (RQ 2). Our analysis of the research gaps and thus our answer to (RQ 3) follows in \autoref{sec:research_gaps}. \autoref{sec:conclusion} concludes this paper.

\section{Related Work}
%!TEX root = ../main.tex
\label{sec:related_work}
Before we begin to collect and analyze the large body of literature on data quality (DQ), we ascertain the extent to which this work has been previously conducted.
In the past, several surveys about DQ definition have been published.
We distinguish between surveys that are based on (1) the result of a systematic literature review and (2) a set of subjectively selected works by the respective authors. Note that we only consider surveys on the definition of DQ, not on the assessment of DQ or other topics related to DQ.

\subsection{Surveys based on Systematic Literature Review (SLR)}
The following surveys conducted a systematic literature review to compare existing definitions of data or information quality.\\*
Wang et al.~\cite{wang1995framework} conducted a systematic review of the literature on DQ up to 1993.
The authors searched for publications that focus on data and information quality.
Among others they asked \emph{how DQ is defined.}
They already observed a lack of consensus on how to define DQ and discussed possible approaches to establishing a definition.
One approach is to use a scientific method, such as an ontological approach, an application of information theory, or an empirical approach based on user studies.
An alternative approach is a pragmatic choice of a definition of DQ that is fit for use, or the establishment of a committee to develop a DQ standard.
Several of these approaches have been taken, such as user studies (e.g., Wang \& Strong~\cite{wang1996}) and the development of a DQ standard~\cite{iso2008} by the International Standardization Organisation (ISO).
Besides this SLR on the definition of DQ in general, there are a few SLRs on \emph{the definition of DQ in specific domains or for specific data representations.}
Zaveri et al.~\cite{zaveri2015} focused on the quality assessment of linked data and asked specifically about the quality dimensions that are relevant in this context.
After systematically collecting the relevant publications and analyzing them, they identified 18 different dimensions of DQ.
These dimensions were grouped into four categories of dimensions: accessibility, intrinsic, contextual, and representational.\\*
Weiskopf et al.~\cite{weiskopf2013} conducted a SLR on the quality assessment of electronic health record data.
They compared the literature along data representation, quality dimensions and quality assessment methods.
Similarly, Liu \& Chi~\cite{liu2002} conducted a SLR on DQ in IoT.
To define DQ in IoT, they basically considered the quality dimensions that play a major role in the publications considered.
Priestley et al.~\cite{priestley2023survey} conducted a SLR on DQ for machine learning (ML).
They also compared the literature along quality dimensions and identified several challenges, namely ethical and legal requirements, data volume, adherence to representational standards, especially metadata, quality of software infrastructures processing the data, and documentation of datasets.

\subsection{Survey Based on Publications Selected by the Authors}
The following surveys are based on a subjective selection of publications chosen by the corresponding authors, i.e. they do not mention a systematic literature review in their publication.\\*

\subsubsection{Definition of quality dimensions for data in general}
The earlier surveys on DQ identified and listed the DQ dimensions they found in the literature and compared definitions.
Wang \& Strong~\cite{wang1996} already noted in their 1996 survey that there is a lack of consensus on the quality dimensions considered.
Scannapieco et al.~\cite{scannapieco2002} reviewed six publications for selected DQ dimensions.
They discussed \emph{the correspondence between different definitions of quality dimensions.}
To do so, they classified the different definitions using the following classification features: approach to dimension definition, modeling view on data, measurement view on data, and context dependency.
The proposed classification aims to guide the designer in selecting DQ dimensions that fit their application needs.
Batini et al.~\cite{batini2009} took a closer look at the definitions of DQ dimensions using essentially the same publications as Scannapieco.
The purpose of their survey was to gain a deeper understanding of the similarities and differences between the defined quality dimensions.
Sidi et al.~\cite{sidi2012} also presented a comparison of different definitions of DQ according to the defined quality dimensions.
In addition to~\cite{scannapieco2002} and \cite{batini2009}, they considered a broader range of publications, encompassing numerous additional quality dimensions.
They collected different definitions for one and the same dimension, but did not compare them in any deeper way.
Laranjeiro et al.~\cite{laranjeiro2015} reviewed the literature on DQ by focusing on the categorization structure and terms used to classify DQ aspects.
As a result of their literature review, Laranjeiro et al. stated that although ``the 'fitness for use' concept is widely accepted, the heterogeneity in the structure and naming and definition of dimensions is very clear``\cite{laranjeiro2015}.
They also identified the \emph{most frequently cited quality dimensions:} accessibility, accuracy, completeness, consistency, and currency.
The motivation behind the survey in Cichy et al.~\cite{cichy2019} was to provide an overview of complete DQ frameworks that are widely applicable.
A fundamental part of the review is a comparison of the underlying DQ definitions.
They also noted that there is a relatively large variation in the set of DQ dimensions considered per framework.
The most common dimensions they mentioned are completeness, accuracy, timeliness, consistency, and accessibility (similar to~\cite{laranjeiro2015}).

\subsubsection{Comparison of data quality definitions}
If quality dimensions are defined so differently and their comparison is difficult therefore, there are basically two ways to compare them anyway: (1) the defined quality dimensions are clustered along coarser criteria leading to clusters of quality dimensions or (2) the differences and commonalities of quality dimensions are identified leading to fine-grained quality requirements or aspects. We first discuss surveys that cluster quality dimensions.

Lacagnina et al.~\cite{lacagnina2023} reviewed several classification mechanisms for DQ definitions and illustrated them with selected publications.
The authors primarily addressed the distinction between quality characteristics and requirements, several clusterings of quality dimensions, the distinction between quality control and quality assurance, and the difference between \emph{fitness for use} and \emph{fitness for purpose}.
Wang et al.~\cite{wang2024} classified the DQ literature in their 2024 survey into three main groups: intuitive, theoretical, and empirical approaches.
Mohammed et al.~\cite{mohammed2024} considered several facets of DQ in order to provide a clustering of quality dimensions.
They also identified the challenge that the definitions of quality dimensions are inherently ambiguous, which can lead to problems in explaining assessment results to consumers.

Instead of clustering quality dimensions, they can also be investigated for differences and commonalities, leading to a more fine-grained view on DQ. The following two surveys took this direction.
Kulikowski et al.~\cite{kulikowski2014} considered a variety of DQ  definitions proposed in the literature.
That paper presents a comprehensive list of \emph{quality attributes} identified in the literature, accompanied by a detailed reason for each.
They conclude that (1) not all DQ attributes can be used to characterize the quality of various types of data or are suitable to the needs of different applications.
(2) Not all DQ attributes are possible or easy to parameterize.
Different parameterized DQ attributes may be expressed in different scales and units.
(3) Not all DQ attributes are independent of each other.

\subsubsection{Definition of data quality for specific domains}
The following literature reviews are dedicated to certain application domains. They basically select suitable literature and identify the quality dimensions that are most important for the domain.
Kandari et al.~\cite{kandari2011} reviewed selected literature on information quality frameworks and attempted to identify the dimensions that are relevant in the context of the World Wide Web.
They followed a two-step process to define DQ: First, finding common dimensions.
Second, to finalize and define the dimensions in the context of the Web.
They subjectively decided that the definitions of Wang \& Strong~\cite{wang1996} were often the most convincing ones.
Schaal et al.~\cite{schaal2012} selected six publications to identify dimensions of information quality for the social web.
First, they identified various relations between quality criteria.
Then, they classified the dimensions found using a semiotic approach, thereby distinguishing between syntax, semantics, and several forms of pragmatics.
Accordingly, Ge et al.~\cite{ge2011} reviewed a number of selected definitions of information quality along a list of quality dimensions.
They identified the most important dimensions as those that are mentioned in at least half of the literature.
Karkouch et al.~\cite{karkouch2016} reviewed quality dimensions for data in general and for Internet of Things (IoT) data from selected literature. Then, they identified common dimensions and interpreted them in the context of IoT.
Zhou et al.~\cite{zhou2024} provided an overview of existing surveys related to DQ in ML, some of which are domain-independent, but most of which are specific to ML.
Specifically, they reconsidered the definition of DQ in Wang \& Strong~\cite{wang1996} and interpreted it in the context of ML.

\autoref{tab:related_surveys} summarizes the surveys discussed above.
It reflects the surveys covered in this section and is not the result of an independent systematic search for surveys.
For each survey, we indicate whether it is based on a systematic literature review, its scope, and the extent to which it classifies DQ definitions.

\begin{table*}
\caption{Comparison of prior surveys on the definition of data quality (DQ).}%
\label{tab:related_surveys}
\begin{tabular}{llll}
    \toprule
    Survey & SLR & Scope & Classification of DQ definitions \\
    \midrule
    \multicolumn{4}{l}{\emph{Surveys based on a systematic literature review}} \\
    Wang et al.~\cite{wang1995framework} & yes & general (until 1993) & none \\
    Zaveri et al.~\cite{zaveri2015} & yes & linked data & dimension clusters (4 categories) \\
    Weiskopf et al.~\cite{weiskopf2013} & yes & healthcare (EHR data) & none \\
    Liu \& Chi~\cite{liu2002} & yes & Internet of Things & none \\
    Priestley et al.~\cite{priestley2023survey} & yes & machine learning & none \\
    \midrule
    \multicolumn{4}{l}{\emph{Surveys based on publications selected by the authors}} \\
    Wang \& Strong~\cite{wang1996} & no & general & none \\
    Scannapieco et al.~\cite{scannapieco2002} & no & general & classification features (4) \\
    Batini et al.~\cite{batini2009} & no & general & none \\
    Sidi et al.~\cite{sidi2012} & no & general & none \\
    Laranjeiro et al.~\cite{laranjeiro2015} & no & general & categorization structure \\
    Cichy et al.~\cite{cichy2019} & no & general & none \\
    Lacagnina et al.~\cite{lacagnina2023} & no & general & classification mechanisms \\
    Wang et al.~\cite{wang2024} & no & general & provenance (3 categories) \\
    Mohammed et al.~\cite{mohammed2024} & no & general & dimension clusters \\
    Kulikowski et al.~\cite{kulikowski2014} & no & general & none \\
    Kandari et al.~\cite{kandari2011} & no & World Wide Web & none \\
    Schaal et al.~\cite{schaal2012} & no & social web & semiotic classification \\
    Ge et al.~\cite{ge2011} & no & general & none \\
    Karkouch et al.~\cite{karkouch2016} & no & Internet of Things & none \\
    Zhou et al.~\cite{zhou2024} & no & ML & none \\
    \midrule
    This survey & yes & general & feature model \\
    \bottomrule
\end{tabular}
\end{table*}

\subsection{Summary} There are only \num{5} systematic literature reviews (SLR) on dimension-based DQ so far (cf.~\autoref{tab:related_surveys}).
The earliest SLR by Wang et al.~\cite{wang1995framework} only includes publications prior to 1993.
The SLR by Zaveri et al.~\cite{zaveri2015} was conducted for a specific type of data.
The SLRs by Weiskopf et al.~\cite{weiskopf2013}, Liu \& Chi~\cite{liu2002} and Priestley et al.~\cite{priestley2023survey} were conducted for DQ in specific domains.
As various definitions of DQ have been published over the last decades, we see \emph{a clear need for an up-to-date systematic literature review on DQ that identifies quality dimensions independent of any data representation and domain.}\\*

Several of the above SLRs on DQ definitions have noted that there is a lack of consensus on quality dimensions.
To provide an overview, the authors have presented clustering approaches for the variety of quality dimensions or identified differences and commonalities of quality dimensions.
Only a few literature reviews have classified the types of data, for which quality has been defined (e.g., Lacagnina et al.~\cite{lacagnina2023}).
Wang et al.~\cite{wang2024} has been the only publication that considers the provenance of quality definitions and classifies them into three categories: intuitive, theoretical, and empirical approaches.
Although classification aspects for DQ definitions appear in several surveys, as \autoref{tab:related_surveys} shows, \emph{none of the existing surveys classifies DQ definitions as a whole.
Existing classification aspects are limited to individual dimensions or specific facets, such as clustering or provenance.}

\section{Systematic Literature Review}
%!TEX root = ../main.tex
\label{sec:slr}

To obtain a comprehensive overview of published data quality (DQ) definitions, we conducted a systematic literature review (SLR) following the methodology proposed by Kitchenham and Charters~\cite{kitchenham2007}.
Our SLR targeted four of the largest academic research platforms, namely ACM Digital Library, Google Scholar, IEEE Xplore Digital Library, and ScienceDirect.
We started by defining the research questions, followed by the search terms and criteria for inclusion and exclusion.
Due to the limitations of the search engines, we extended the SLR using the snowballing approach of Wohlin~\cite{wohlin2014}, starting with the six most influential papers found in the SLR.
Finally, we added the acknowledged ISO standard for DQ.
The identification and selection process is summarized in the PRISMA 2020 flow diagram~\cite{page2021} in \autoref{fig:prisma} and described in detail below.
We ended up with \num{35} relevant publications on DQ definitions.

\begin{figure}
  \centering
  \definecolor{prismablue}{RGB}{189,215,238}
  \definecolor{prismablueL}{RGB}{228,239,249}
  \resizebox{\textwidth}{!}{%
  \begin{tikzpicture}[
      node distance=5mm and 4mm,
      box/.style={draw, text width=44mm, align=left, font=\small,
                  inner sep=1.5mm},
      sidebox/.style={box, text width=34mm},
      idbox/.style={box, minimum height=34mm},
      heading/.style={draw, fill=prismablue, align=center,
                      font=\small\bfseries, inner sep=1.5mm,
                      minimum height=8mm, text width=82mm},
      phase/.style={draw, fill=prismablueL, align=center,
                    font=\small\bfseries, inner sep=1.5mm,
                    rotate=90, minimum height=7mm},
      arr/.style={-{Stealth[length=2mm]}, semithick}
    ]

    % --- identification via databases ---
    \node[idbox] (db-id) {Records identified from databases (searched July~2024) (n~=~7{,}170):\\
      \quad ACM DL (n~=~835)\\
      \quad Google Scholar (n~=~1{,}000)\textsuperscript{c}\\
      \quad IEEE Xplore (n~=~2{,}209)\\
      \quad ScienceDirect (n~=~3{,}126)};
    \node[box, below=8mm of db-id] (db-scr)
      {Records screened (title and abstract) (n~=~7{,}170)};
    \node[sidebox, right=of db-scr] (db-scr-ex)
      {Records excluded (n~=~6{,}825)};
    \node[box, below=of db-scr] (db-ret)
      {Reports sought for retrieval (n~=~345)};
    \node[sidebox, right=of db-ret] (db-ret-ex)
      {Reports not retrieved (n~=~2)};
    \node[box, below=of db-ret, minimum height=27mm] (db-eli)
      {Reports assessed for eligibility (full text) (n~=~343)};
    \node[sidebox, right=of db-eli, minimum height=27mm] (db-eli-ex)
      {Reports excluded (n~=~323):\\
      \quad Duplicates (n~=~8)\\
      \quad EC1 (n~=~258)\\
      \quad EC2 (n~=~18)\\
      \quad EC3 (n~=~39)};

    % --- identification via other methods ---
    \node[idbox, right=46mm of db-id] (ot-id) {Records identified from:\\
      \quad Citation searching:\\
      \quad forward snowballing\textsuperscript{a}\\
      \quad (n~=~1{,}539), backward\\
      \quad snowballing\textsuperscript{b} (n~=~12)\\
      \quad Standards: ISO~25012 (n~=~1)};
    \node[box] (ot-ret) at (ot-id |- db-ret)
      {Reports sought for retrieval (n~=~42)};
    \node[box, minimum height=27mm] (ot-eli) at (ot-id |- db-eli)
      {Reports assessed for eligibility (n~=~42)};
    \node[sidebox, right=of ot-eli, minimum height=27mm] (ot-eli-ex)
      {Reports excluded (n~=~27):\\
      \quad EC1--EC3 (n~=~26)\\
      \quad Included via database\\
      \quad search (n~=~1)};

    \node[box, below=8mm of db-eli, minimum height=27mm] (inc)
      {Studies included in review (n~=~35):\\
      \quad via databases (n~=~20)\\
      \quad via citation searching (n~=~14)\\
      \quad via standards (n~=~1)};

    \node[heading, anchor=south] (h-db)
      at ($(db-id.north west)!0.5!(db-ret-ex.north east |- db-id.north east)+(0,4mm)$)
      {Identification of studies via databases};
    \node[heading, anchor=south]
      at ($(ot-id.north west)!0.5!(ot-eli-ex.north east |- ot-id.north east)+(0,4mm)$)
      {Identification of studies via other methods};

    \node[phase, anchor=south, minimum width=30mm]
      at ($(db-id.west)+(-4mm,0)$) {Identification};
    \node[phase, anchor=south, minimum width=44mm]
      at ($(db-scr.north west)!0.5!(db-eli.south west)+(-4mm,0)$) {Screening};
    \node[phase, anchor=south, minimum width=20mm]
      at ($(inc.west)+(-4mm,0)$) {Included};

    \draw[arr] (db-id) -- (db-scr);
    \draw[arr] (db-scr) -- (db-scr-ex);
    \draw[arr] (db-scr) -- (db-ret);
    \draw[arr] (db-ret) -- (db-ret-ex);
    \draw[arr] (db-ret) -- (db-eli);
    \draw[arr] (db-eli) -- (db-eli-ex);
    \draw[arr] (db-eli) -- (inc);
    \draw[arr] (ot-id) -- (ot-ret);
    \draw[arr] (ot-ret) -- (ot-eli);
    \draw[arr] (ot-eli) -- (ot-eli-ex);
    \draw[arr] (ot-eli.south) -- (ot-eli.south |- inc.east) -- (inc.east);
  \end{tikzpicture}}
  \caption{PRISMA 2020 flow diagram~\cite{page2021} of the study
    identification and selection process of our systematic literature
    review. EC1--EC3 denote the exclusion criteria defined in
    \autoref{sec:slr_review_methodology}.
    \textsuperscript{a}\,First iteration only.
    \textsuperscript{b}\,Reference lists of excluded surveys.
    \textsuperscript{c}\,First 1{,}000 of ${\sim}$10{,}500 hits retrievable.}%
  \label{fig:prisma}
  \Description{PRISMA 2020 flow diagram showing the number of records
    identified via database search, citation searching, and standards,
    the records excluded during screening and eligibility assessment,
    and the \num{35} studies included in the review.}
\end{figure}

\subsection{Systematic Literature Review Methodology}%
\label{sec:slr_review_methodology}

Following the approach outlined by Kitchenham and Charters~\cite{kitchenham2007}, the first step is to define the research question, followed by the formulation of a search strategy and then the selection of publications.

\paragraph{Research question.}
The aim of this survey is to evaluate existing definitions of DQ.
Since DQ is a multifaceted concept that is typically defined in terms of quality dimensions, we look specifically for DQ definitions that mention quality dimensions.
In the context of DQ, the term \emph{dimension} is used to describe fundamental, qualitative characteristics that represent different aspects of the fitness of data for its intended purpose.
A set of these dimensions, such as completeness, accuracy, consistency, and timeliness, serve as a conceptual framework for defining DQ.
Dimensions are inherently qualitative and provide a basis for understanding and assessing DQ at a conceptual level.
While a dimension defines the aspect of quality being evaluated, a metric serves as a measurable, quantitative expression that supplements the dimension to form the next step in a quality assurance process.
Therefore, we exclude publications that do not define DQ dimensions but \emph{only} metrics.
However, we include definitions regardless of whether they are entirely new definitions or merely based on existing definitions from other publications.
The only exceptions are definitions that clearly refer to an existing work.
In such cases, only the original work is considered.
In conclusion, we address (RQ 1), which was defined in \autoref{sec:introduction}:

\begin{enumerate}
    \item[(RQ 1)] Which publications contain original, dimension-based definitions of data quality?
\end{enumerate}

\paragraph{Search strategy.}
Search strategies in a systematic review are usually iterative and are designed to ensure that all relevant publications are covered.
The following inclusion and exclusion criteria were defined to answer the research questions and to limit the results:
Thus, we define the following inclusion criteria (IC):

\begin{enumerate}
    \item[(IC1)] Publications that assess the quality of data or information in any of the following forms: assessment, assurance, definition, survey, analysis, dimension, framework.
    \item[(IC2)] Publications that are written in English.
\end{enumerate}

\emph{Search terms}.
In order to ensure a very comprehensive search, we define two term groups and the term \emph{quality}.
Given that the term \emph{information} is often used as a synonym for \emph{data}, both terms were included in the systematic search to ensure a broader coverage forming the term group \emph{data type} (TG1).
Since the term \emph{definition} may not be explicitly mentioned, but quality definition is nevertheless included in the context of, for example, quality assessment frameworks or surveys, we define the term group \emph{process type} (TG2).
In conclusion, we define the following terms and groups:

\begin{enumerate}
    \item[(TG1)] \emph{data type} = ["data", "information"]
    \item[(TG2)] \emph{process type} = ["assessment", "assurance", "definition", "survey", "analysis", "dimension", "framework"]
\end{enumerate}

These terms were used as follows:
[data type] $\wedge$ "quality" $\wedge$ [process type]\\
\hfill\break
According to Kitchenham~\cite{kitchenham2007}, a title-only search does not always return enough relevant publications.
Since a full-text search leads to many irrelevant publications, we decided to focus the query primarily on the title for \emph{data type} (TG1) and \emph{quality}, and secondarily on the abstract for the term group \emph{process type} (TG2).
This corresponds to our research questions when searching for specific publications.
Data and information quality is the key to our search, so we expected \emph{data type} (TG1) and \emph{quality} in the title.
There is no clear terminology for \emph{process type} (TG2), so we expected the terms of this group not necessarily in the title but in the abstract.

We used the list of search engines and digital libraries listed in \autoref{tab:slr_search} to cover the major databases and publications of the major conferences and journals in computer science.
Since each digital library offers different search functionalities, we chose the search engine-specific settings that came closest to our original search goal.
The table documents the variations for each search performed along with the final search expression used, which is already formatted according to the syntactic guidelines of each search engine.
Note that Google Scholar was only searched for publication titles, as searching in abstracts is not supported.
Altogether, the searches returned \num{16670} hits.
Google Scholar accounted for approximately \num{10500} of them but only permits access to the first \num{1000} results, so only these \num{1000} most relevant hits were considered.
Following the PRISMA guidance for sources with retrieval limits, \autoref{fig:prisma} therefore reports the \num{1000} retrievable records for Google Scholar.
All results from the other search engines were reviewed in full, which leads to a total of \num{7170} records screened by title and abstract.
To increase the likelihood of finding all relevant publications, this approach was complemented by an exhaustive snowballing process (cf.~\autoref{sec:slr_snowballing}).

\begin{table*}
  \begin{threeparttable}

  \caption{Search strategy per source: scope and search expression}%
  \label{tab:slr_search}

  \begin{tabular}{@{}>{\raggedright\arraybackslash}p{0.16\columnwidth}>{\raggedright\arraybackslash}p{0.2\columnwidth}>{\raggedright\arraybackslash\hyphenpenalty=10000\exhyphenpenalty=10000}p{0.57\columnwidth}@{}}
    \toprule
    Source & Scope \& Restriction & Search expression\\
    \midrule

    ACM Digital Library\tnote{1} & title \& abstract & Title:(data OR information) AND Title:(quality) AND Abstract:(assess* OR assur* OR defin* OR analy* OR "dimension" OR "framework")\\

    Google Scholar\tnote{2} & title \& en & allintitle:((information OR data) AND quality AND (assessment OR assurance OR definition OR survey OR analysis OR dimension OR framework)) \\

    IEEE Xplore Digital Library\tnote{3} & title \& abstract & ("Document Title":"information" OR "Document Title":"data") AND "Document Title":"quality" AND ("Abstract":assess* OR "Abstract":assur* OR "Abstract":defin* OR "Abstract":analy* OR "Abstract":"dimension" OR "Abstract":"framework") \\

    Science Direct\tnote{4} & title \& abstract & Title:(data AND quality) AND Abstract:(assessment OR assurance OR definition OR analysis OR dimension OR framework) \\
    \bottomrule
  \end{tabular}
      \begin{tablenotes}
        \item[1] \url{https://dl.acm.org/search/advanced}
        \item[2] \url{https://scholar.google.com}
        \item[3] \url{https://ieeexplore.ieee.org/search/advanced/command}
        \item[4] \url{https://www.sciencedirect.com/search/entry}
      \end{tablenotes}
  \end{threeparttable}
\end{table*}

For each search result, we evaluated the title and abstract to determine whether a publication actually defined DQ dimensions.
In cases where the title and abstract were not sufficiently transparent and the characteristics or dimensions of DQ were not readily apparent, we included these publications.
The full text of all resulting potential publications was reviewed.
Of the \num{7170} records screened, \num{345} reports were sought for retrieval.
Two of them could not be retrieved, so \num{343} reports were assessed for eligibility based on their full text.
\paragraph{Selection of literature.}
The hits resulting from the previous step were checked against the defined exclusion criteria~(EC):

\begin{enumerate}
    \item[(EC1)] Publications that do not define at least one DQ dimension.
    \item[(EC2)] Publications that are not published as articles in peer-reviewed journals or conference proceedings, or that are not published by organizations concerned with data management or standardization issues.
    \item[(EC3)] Publications that are preprints or short versions of other publications and therefore do not represent a new contribution.
\end{enumerate}

In the case of publications that merely cite the dimensions of other publications (EC3), we applied backward snowballing to identify the originals and repeated the process.
In this way, we retrieved \num{12} reports from the reference lists of the excluded publications.
The bibliographic metadata of the resulting relevant publications were then captured using the bibliography management platform Zotero\footnote{\url{https://www.zotero.org}}, where \num{8} duplicate records across the four sources were identified and removed.
Of the \num{343} reports assessed, we excluded \num{323} in total, namely \num{8} duplicates, \num{258} publications according to EC1, \num{18} according to EC2, and \num{39} according to EC3.
In the end, we identified \num{20} publications directly via the database searches.
The \num{8} original publications found through backward snowballing of excluded reports are counted as identified via citation searching in \autoref{fig:prisma}.
The screening and eligibility assessment were performed by the first author.
The third author regularly checked whether community-driven publications were included in the set under review.
In cases where there was uncertainty about the inclusion or exclusion of a particular publication, all authors participated in a collective discussion that resulted in a consensus on the candidate.
Finally, all authors conducted a final review of the resulting list of 35 publications to confirm that the exclusion criteria had been met.

\subsection{Forward Snowballing}%
\label{sec:slr_snowballing}

According to Kitchenham~\cite{kitchenham2007}, the goal of an SLR is to identify all relevant publications.
However, the search methodology is no more effective than the search string used.
As Wohlin~\cite{wohlin2014} stated, formulating effective search strings is a challenging endeavor, especially given the lack of standardization of terminology.
He added methods of snowballing to existing guidelines for systematic literature reviews.
Wohlin's guideline combines backward and forward snowballing with a systematic approach to identifying the references and citations of relevant publications.
Since backward snowballing had already been applied during the selection of literature (cf.~\autoref{sec:slr_review_methodology}), we limited this step to forward snowballing.
Furthermore, our conventional literature search covers the literature that has been around for some time.

The initial phase of the snowballing process involved identifying a starter set of publications.
According to Wohlin, a comprehensive starter set should contain publications from different communities, publishers, years, and authors to ensure a broad and diverse representation.
In addition to these criteria, we favored publications with a high impact factor in the starter set.
The importance of snowballing and the composition of the starter set were discussed and agreed upon among all authors, resulting in the selection of the following six publications:
Wang \& Strong~\cite{wang1996}, Wand \& Wang~\cite{wand1996}, Batini et al.~\cite{batini2009}, Kahn et al.~\cite{kahn2016}, Zaveri et al.~\cite{zaveri2015}, and Ballou \& Pazer~\cite{ballou1985}. 
Ballou \& Pazer~\cite{ballou1985}, Wang \& Strong~\cite{wang1996}, Wand \& Wang~\cite{wand1996}, Batini et al.~\cite{batini2009} are publications that contain quality definitions for data in general, while Zaveri et al.~\cite{zaveri2015} and Kahn et al.~\cite{kahn2016} define quality for a specific type of data (i.e., linked data) and in a specific domain (i.e., healthcare).
To avoid inadvertently excluding publications that use a particular terminology, we chose a broader search term than the one used in \autoref{sec:slr_review_methodology} and removed the term group \emph{process type}.
Accordingly, we decided to use predefined terms as follows:
(“data” $\vee$ "information") $\wedge$ "quality" \\*
\hfill\break
According to Wohlin~\cite{wohlin2014}, we applied the forward snowballing approach manually using the bibliography of each publication with Google Scholar as the search engine, since
Google Scholar supports searches for forward citations very conveniently.
New publications are subject to the same exclusion criteria as the literature selection.
In accordance with the guidelines, we used the snowballing method until no new publications were added.
The first iteration alone yielded \num{1539} citing records for the six starter publications (see \autoref{fig:prisma}).
The sizes of the subsequent iterations were not recorded.
Publications not already included in the Zotero list mentioned above were added to the list.
We ended up with \num{6} additional publications using this approach.
Together with the \num{8} publications from backward snowballing, this amounts to the \num{14} publications identified via citation searching in \autoref{fig:prisma}.

\subsection{The ISO Standard}
The methodologies for SLRs are well suited for scientific publications but have limitations for documents published by organizations for standardization such as the International Organization for Standardization (ISO).
The ISO standard 25012~\cite{iso2008}
cannot be found using the search engines. As this is the quintessential definition of DQ that has led to an official standard, we manually included it in our analysis.

\subsection{Results}%
\label{sec:slr_results}

As a result of this systematic search, we identified \num{35} relevant publications, which are listed in \autoref{tab:slr_publications} and available as a Zotero public library\footnote{\url{https://www.zotero.org/groups/6741534/data_quality_definitions_literature_survey}}.
The publications in this list make a novel contribution by defining new dimensions or adapting existing ones to specific domains or datasets.
Publications that merely cite the quality dimensions defined in other publications are not included in this list.
From the large number of publications on data and information quality that we initially found in our literature search, \num{35} publications is a very small percentage.
Potential explanations for this small number are that the publications frequently refer to existing DQ definitions or do not consider the definition of DQ important enough to make it explicit.
This is consistent with our deliberately recall-oriented search strategy (cf.~\autoref{sec:slr_review_methodology}), which matches the broad term group \emph{process type} against the abstract to avoid missing relevant publications at the cost of retrieving many that merely mention DQ.
Indeed, the exclusion counts in \autoref{fig:prisma} confirm this: EC1 alone accounts for \num{258} of the \num{323} excluded reports, i.e., the vast majority of retrieved publications use the term \emph{data quality} without originally defining it.

As DQ can play an important role not only in computer science but also in several other fields, we investigated the venues of the listed articles using Semantic Scholar\footnote{\url{https://www.semanticscholar.org}}.
A majority of the publications were published in computer science.
\num{13} publications are related to computer science and another field.
Only \num{5} publications were not published in computer science.
The following other fields were selected for publication:
\num{8} publications are related to business, \num{4} to medicine, \num{3} to engineering, \num{2} to environmental science, \num{1} to sociology, and \num{1} to political science.
These figures show that \emph{DQ is important not only in computer science, but also in other areas.}

There are several \emph{highly cited publications}, the top 8 publications according to Google Scholar are Wang \& Strong~\cite{wang1996}, Wand \& Wang~\cite{wand1996}, Strong et al.~\cite{strong1997}, Batini et al.~\cite{batini2009}, Cai et al.~\cite{cai2015}, Ballou \& Pazer~\cite{ballou1985}, Zaveri et al.~\cite{zaveri2015}, and Kahn et al.~\cite{kahn2016}.
The high citation numbers show that the community uses a relatively small number of publications to refer to a definition of DQ.
We analyze these and the other 27 selected publications to find out why they are so highly cited.
Note that Google Scholar does not provide citation numbers for ISO 25012~\cite{iso2008} and \cite{dama2013} by the DAMA UK WG.
All the publications in \autoref{tab:slr_publications} form the basis of our feature-oriented domain analysis in \autoref{sec:foda}.

\begin{table*}
    \caption{List of publications that define data quality (DQ) dimensions}%
    \label{tab:slr_publications}
    \scriptsize
    \begin{tabular}{l>{\raggedright\arraybackslash}p{2.2cm}p{6.2cm}>{\raggedright\arraybackslash}p{2.8cm}r}
        \toprule
        Year & Authors & Title & Field & \#Cit.\\
        \midrule
        1985 & Ballou \& Pazer~\cite{ballou1985} & Modeling Data and Process Quality in Multi-Input, Multi-Output Information Systems & Computer Science & 939\\
        1996 & Wang \& Strong~\cite{wang1996} & Beyond Accuracy: What Data Quality Means to Data Consumers & Computer Science, Business & 6921\\
        1996 & Wand \& Wang~\cite{wand1996} & Anchoring data quality dimensions in ontological foundations & Computer Science, Business & 2352\\
        1997 & Strong et al.~\cite{strong1997} & Data Quality in Context & Computer Science & 2131\\
        1999 & Jarke et al.~\cite{jarke1999} & Architecture and quality in data warehouses: An extended repository approach & Computer Science & 315\\
        1999 & Shanks \& Corbitt~\cite{shanks1999} & Understanding data quality: Social and cultural aspects & Sociology, Computer Science & 148\\
        1999 & Brackstone~\cite{brackstone1999} & Managing data quality in a statistical agency & Business & 224\\
        2002 & Liu \& Chi~\cite{liu2002} & Evolutional Data Quality: A Theory-Specific View & Computer Science & 170\\
        2004 & Scannapieco et al.~\cite{scannapieco2004} & The DaQuinCIS architecture: a platform for exchanging and improving data quality in cooperative information systems & Computer Science & 177\\
        2006 & Verma~\cite{verma2006} & Data quality and clinical audit & Computer Science, Medicine & 23\\
        2007 & Caro et al.~\cite{caro2007} & A Probabilistic Approach to Web Portal's Data Quality Evaluation & Computer Science & 22\\
        2007 & Reddy et al.~\cite{reddy2007} & A framework for data quality and feedback in participatory sensing & Computer Science, Engineering & 47\\
        2007 & Klein et al.~\cite{klein2007streaming} & Representing Data Quality for Streaming and Static Data & Computer Science, Engineering & 64\\
        2007 & Klein~\cite{klein2007sensor} & Incorporating Quality Aspects in Sensor Data Streams & Computer Science, Business & 51\\
        2008 & ISO 25012~\cite{iso2008} & Software engineering — Software product Quality Requirements and Evaluation (SQuaRE) — Data quality model & Computer Science & -\\
        2009 & Batini et al.~\cite{batini2009} & Methodologies for data quality assessment and improvement & Computer Science & 1886\\
        2011 & Kandari et al.~\cite{kandari2011} & Information quality on the World Wide Web: development of a framework & Computer Science & 27\\
        2011 & Ge et al.~\cite{ge2011} & Information quality assessment: Validating measurement dimensions and processes & Computer Science & 53\\
        2011 & Chiasera et al.~\cite{chiasera2011} & Federated EHR: How to improve data quality maintaining privacy & Computer Science, Medicine & 4\\
        2011 & Friberg et al.~\cite{friberg2011} & Information quality criteria and their importance for experts in crisis situations & Political Science & 24\\
        2012 & Schaal et al.~\cite{schaal2012} & Information Quality Dimensions for the Social Web & Business, Computer Science & 26\\
        2013 & Montori et al.~\cite{montori2013} & Basing information on comprehensive, critically appraised, and up-to-date syntheses of the scientific evidence: a quality dimension of the International Patient Decision Aid Standards & Medicine & 96\\
        2013 & DAMA UK WG~\cite{dama2013} & The six primary dimensions for data quality assessment & Business & -\\
        2014 & Kulikowski~\cite{kulikowski2014} & Data Quality Assessment: Problems and Methods & Computer Science & 9\\
        2015 & Zaveri et al.~\cite{zaveri2015} & Quality assessment for Linked Data: A Survey: A systematic literature review and conceptual framework & Computer Science & 670\\
        2015 & Cai \& Zhu~\cite{cai2015} & The Challenges of Data Quality and Data Quality Assessment in the Big Data Era & Computer Science & 1421\\
        2016 & Kahn et al.~\cite{kahn2016} & A Harmonized Data Quality Assessment Terminology and Framework for the Secondary Use of Electronic Health Record Data & Medicine, Computer Science & 477\\
        2019 & Ceravolo \& Bellini~\cite{ceravolo2019} & Towards Configurable Composite Data Quality Assessment & Computer Science & 12\\
        2019 & El Alaoui et al.~\cite{elalaoui2019} & Big Data Quality Metrics for Sentiment Analysis Approaches & Computer Science & 43\\
        2020 & Firmani et al.~\cite{firmani2020} & Ethical Dimensions for Data Quality & Computer Science & 2\\
        2020 & Azeroual \& Lewoniewski~\cite{azeroual2020}& How to Inspect and Measure Data Quality about Scientific Publications: Use Case of Wikipedia and CRIS Databases & Computer Science, Environmental Science & 10\\
        2020 & Black \& Nederpelt~\cite{black2020} & How to Select the Right Dimensions of Data Quality & Computer Science, Business & 18\\
        2021 & DeCastro-García \& Pinto~\cite{decastro2021} & A Data Quality Assessment Model and Its Application to Cybersecurity Data Sources & Computer Science, Engineering & 2\\
        2023 & Pansara~\cite{pansara2023} & Cultivating Data Quality to Strategies, Challenges, and Impact on Decision-Making & - & 71\\
        2024 & Gao et al.~\cite{gao2024} & Low-carbon information quality dimensions and random forest algorithm evaluation model in digital marketing & Business, Environmental Science & 0\\
        \bottomrule
    \end{tabular}
\end{table*}

\subsection{Limitations}%
\label{sec:slr_limitations}

Like any SLR, our review is subject to limitations.
We summarize them here together with the measures taken to mitigate them.

\paragraph{Distinction between dimensions and metrics.}
The distinction between qualitative dimensions and quantitative metrics (cf.~\autoref{sec:slr_review_methodology}) involves judgment in borderline cases.
Such cases were resolved by consensus among all authors, following the same procedure as for study selection.
Publications that define only metrics do not define a dimension and thus lie outside our research question; excluding them does not affect the completeness of the surveyed dimension-based definitions.

\paragraph{Language and scope restrictions.}
We only considered publications written in English (IC2), so definitions published in other languages may be missing.
Furthermore, our research question deliberately focuses on dimension-based DQ definitions.
Literature from adjacent fields such as data governance, policy, or AI ethics rarely defines DQ in terms of dimensions and is therefore outside the scope of this review.
Consequently, we do not claim completeness across all disciplines.

\paragraph{Terminology bias.}
Our search string requires the terms of TG1 and \emph{quality} to appear in the title, so relevant publications using different terminology in their titles may have been missed.
This is a deliberate trade-off: a full-text search returns a large number of irrelevant results~\cite{kitchenham2007}, which is why only the seven terms of TG2 were matched against the abstract.
To compensate, the snowballing process used a broader search string without the term group \emph{process type} (TG2) and thus also captures publications with divergent terminology, provided they are connected to the citation network of the starter set.

\paragraph{Google Scholar retrieval limit.}
As described in \autoref{sec:slr_review_methodology}, Google Scholar returned approximately \num{10500} hits, of which only the first \num{1000} could be retrieved.
Since Google Scholar ranks results by relevance, the retrieved records are those most likely to be eligible.
Moreover, Google Scholar is only one of four sources, and the snowballing process provides an additional, database-independent path to relevant publications.

\paragraph{Recognizability of dimension definitions.}
Applying exclusion criterion EC1 requires judging whether a publication defines at least one DQ dimension.
We based this judgment on content rather than terminology: publications were included regardless of whether they use the term \emph{dimension}, including those that define only a single dimension or characterize DQ through sets of quality attributes, which is reflected by the definition type \emph{set of attributes} in our feature model (cf.~\autoref{sec:foda}).
Nevertheless, publications whose understanding of DQ is not recognizable from the full text may have been excluded.

\paragraph{Researcher bias in study selection.}
Applying the inclusion and exclusion criteria involves judgment, and screening was primarily carried out by a single author, which poses a risk of selection bias.
This risk was mitigated by the cross-checks, consensus discussions, and the final review by all authors described in \autoref{sec:slr_review_methodology}.

\section{Feature-Oriented Domain Analysis}
%!TEX root = ../main.tex
\label{sec:foda}

Feature-Oriented Domain Analysis (FODA) by Kang et al.~\cite{kang1990} was originally designed to identify the common and variable features of a family of related software systems and to represent them in a feature model.
Features represent distinguishable characteristics of systems in a domain~\cite{kang1990}.
They also have relationships such as mandatory, optional, and alternative relationships, as well as dependencies between features. 
Domain analysis is not limited to software systems but can also be applied to concepts~\cite{czarnecki2006}.
For example, feature models have been used to classify model transformation approaches~\cite{czarnecki2006} and, based on a systematic literature review, model repair approaches~\cite{macedo2017}.

In this study, we adapt FODA to classify the data quality (DQ) definitions identified through our systematic literature review (SLR). 
We use the scope established by the SLR (see \autoref{sec:slr}) as the basis for our domain analysis. 
During domain modeling, we identify the commonalities and variabilities among the DQ definitions and represent them as features and relationships in a feature model. 
We then use this model to classify the publications containing the identified definitions based on the feature configurations of those definitions.

This classification serves two purposes. 
First, it helps us understand the dimensions along which DQ definitions vary. 
Second, it supports the identification of research gaps by highlighting feature configurations that are not represented in the reviewed literature but may warrant further investigation, as discussed in \autoref{sec:research_gaps}. 
For practitioners, the resulting classification provides a structured way to identify DQ definitions that are appropriate for their data types, domains of interest, and contexts of use.

We constructed the feature model iteratively, combining the conceptual-to-empirical and the empirical-to-conceptual approach of Nickerson et al.~\cite{nickerson2013method}.
We started with a set of conceptually derived candidate features reflecting the research questions (RQ2.1)--(RQ2.4).
In each iteration, the publications resulting from the SLR were classified against the current version of the feature model.
Whenever a DQ definition could not be represented adequately, the feature model was revised by adding, splitting, or generalizing features.
All authors contributed to this process, and disagreements were resolved through discussion until consensus was reached.
The process terminated when all \num{35} publications could be classified and an additional pass over the publications yielded no more changes to the feature model, while the requirements stated above were met.
As a result, we present the final feature model in \autoref{sec:foda_feature_model}.
The classification results are presented in \autoref{sec:foda_classification}.

\subsection{Feature Model}%
\label{sec:foda_feature_model}

\begin{figure}[h]
  \centering
  \includegraphics[width=341.7pt]{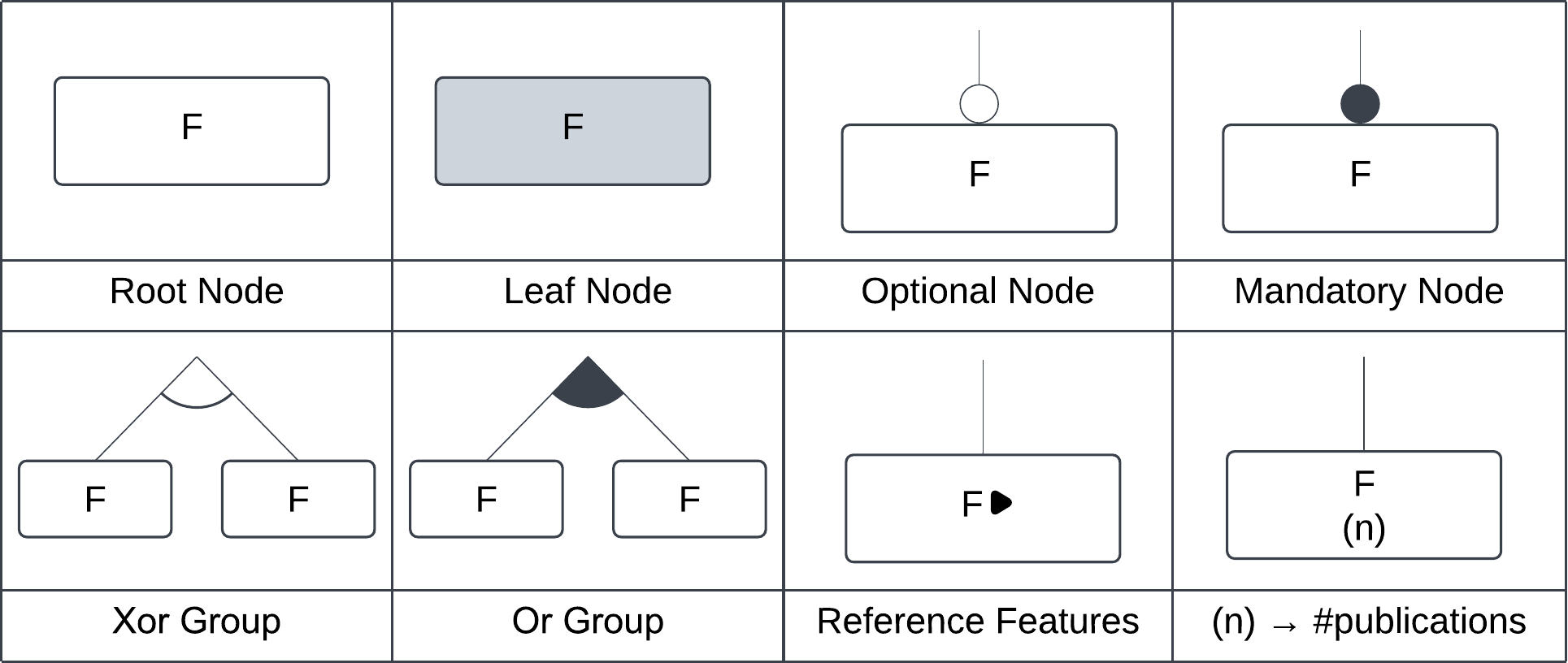}
  \caption{Legend of the diagram components used in the FODA feature model for DQ definitions.}%
  \label{fig:foda_legend}
  \Description{The diagram components of the FODA feature model presented in this paper.}
\end{figure}

We use the feature diagram notation shown in \autoref{fig:foda_legend}, following Macedo et al.~\cite{macedo2017}.
A child feature can only be selected for a publication if its parent is also selected.
Child features can be either \emph{mandatory} or \emph{optional}, or they can be arranged in groups.
In an \emph{xor group}, exactly one feature of the group must be selected if the parent is selected.
In an \emph{or group}, at least one feature of the group must be selected if the parent is selected.
Xor groups are part of the original FODA notation, where they are called alternative features~\cite{kang1990}, while or groups were introduced by Czarnecki and Eisenecker~\cite{czarnecki2000}.
Each feature model has a \emph{root feature}, which is always present in every configuration, and may contain \emph{reference features}, which simply point to other feature models.
We extend this notation by shading leaf features and by annotating each feature with the number (n) of publications that have the feature, as shown in the following diagrams.
The number is not given if exactly all publications have the feature.

In the light of all the publications defining DQ, we present a newly developed feature model for their classification.
Its structure is derived from the following research questions.

\begin{enumerate}
    \item[(RQ2)] How can the data quality definitions found in the literature be classified?
    \begin{enumerate}
        \item[(RQ2.1)] Is data quality defined for a particular type of data?
        \item[(RQ2.2)] What contexts for data are considered when defining data quality?
        \item[(RQ2.3)] How is the definition of data quality given?
        \item[(RQ2.4)] How was the definition of data quality derived from existing information?
    \end{enumerate}
\end{enumerate}

\begin{figure}[h]
  \centering
  \includegraphics[width=\linewidth]{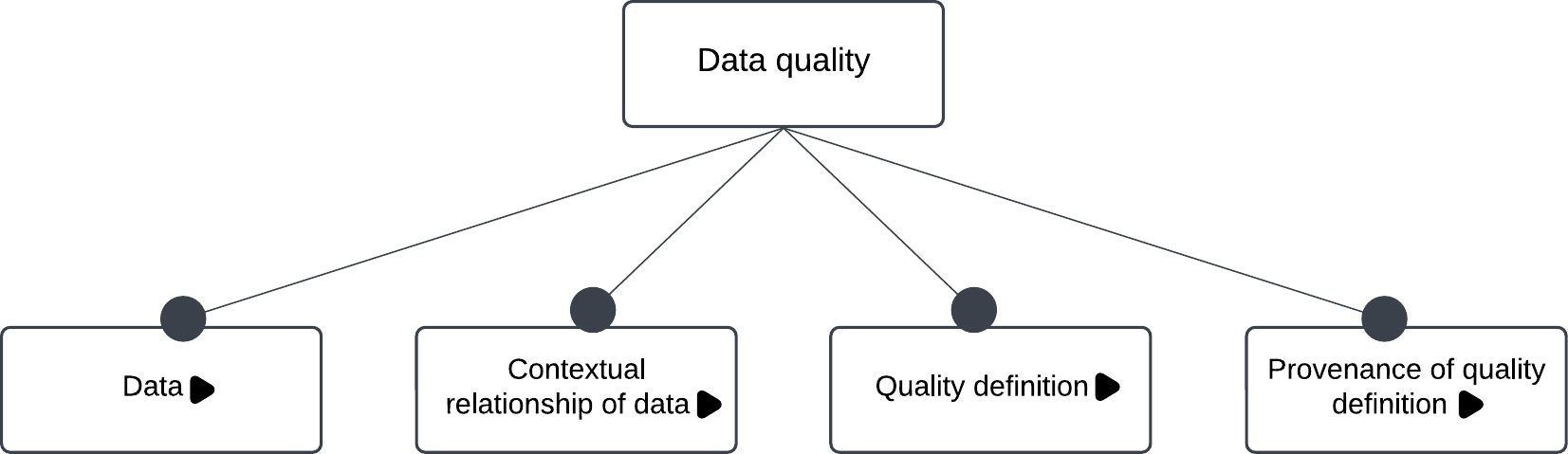}
  \caption{Root feature of the FODA feature model for DQ definitions (legend: cf.~\autoref{fig:foda_legend}).}
  \label{fig:foda_model_root}
  \Description{The root feature of the FODA feature model presented in this paper.}
\end{figure}

\begin{figure}[h]
  \centering
  \includegraphics[width=\linewidth]{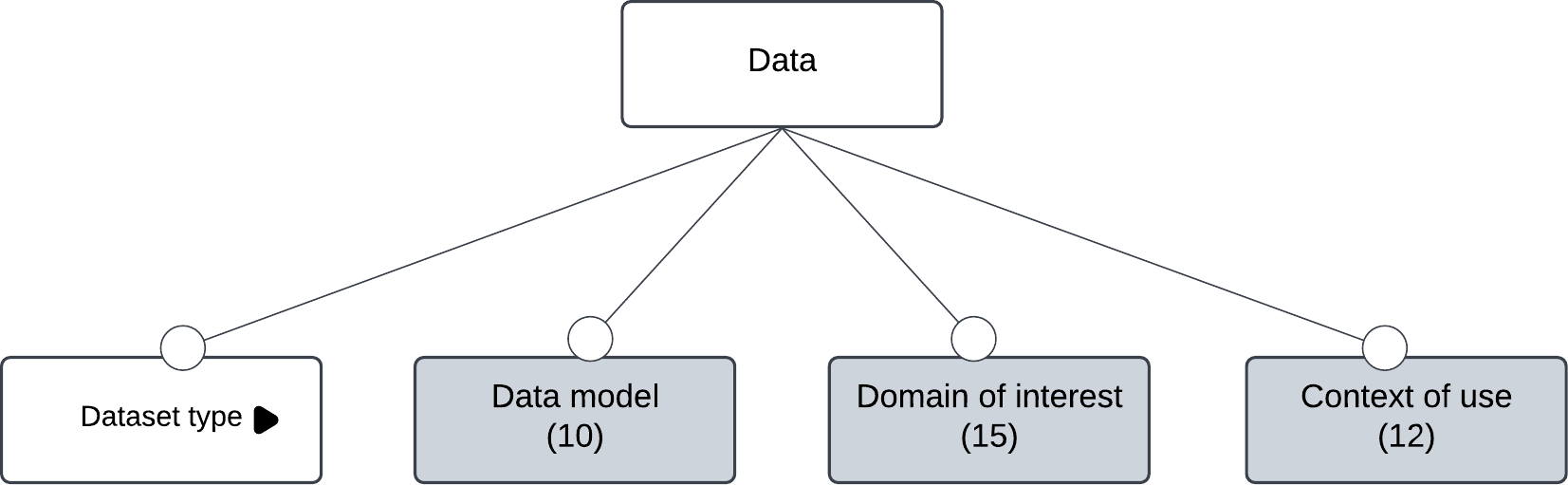}
  \caption{Data feature of the FODA feature model for DQ definitions (legend: cf.~\autoref{fig:foda_legend}).}
  \label{fig:foda_model_data}
  \Description{The 'data' feature of the FODA feature model presented in this paper.}
\end{figure}

\begin{figure}[h]
  \centering
  \includegraphics[width=\linewidth]{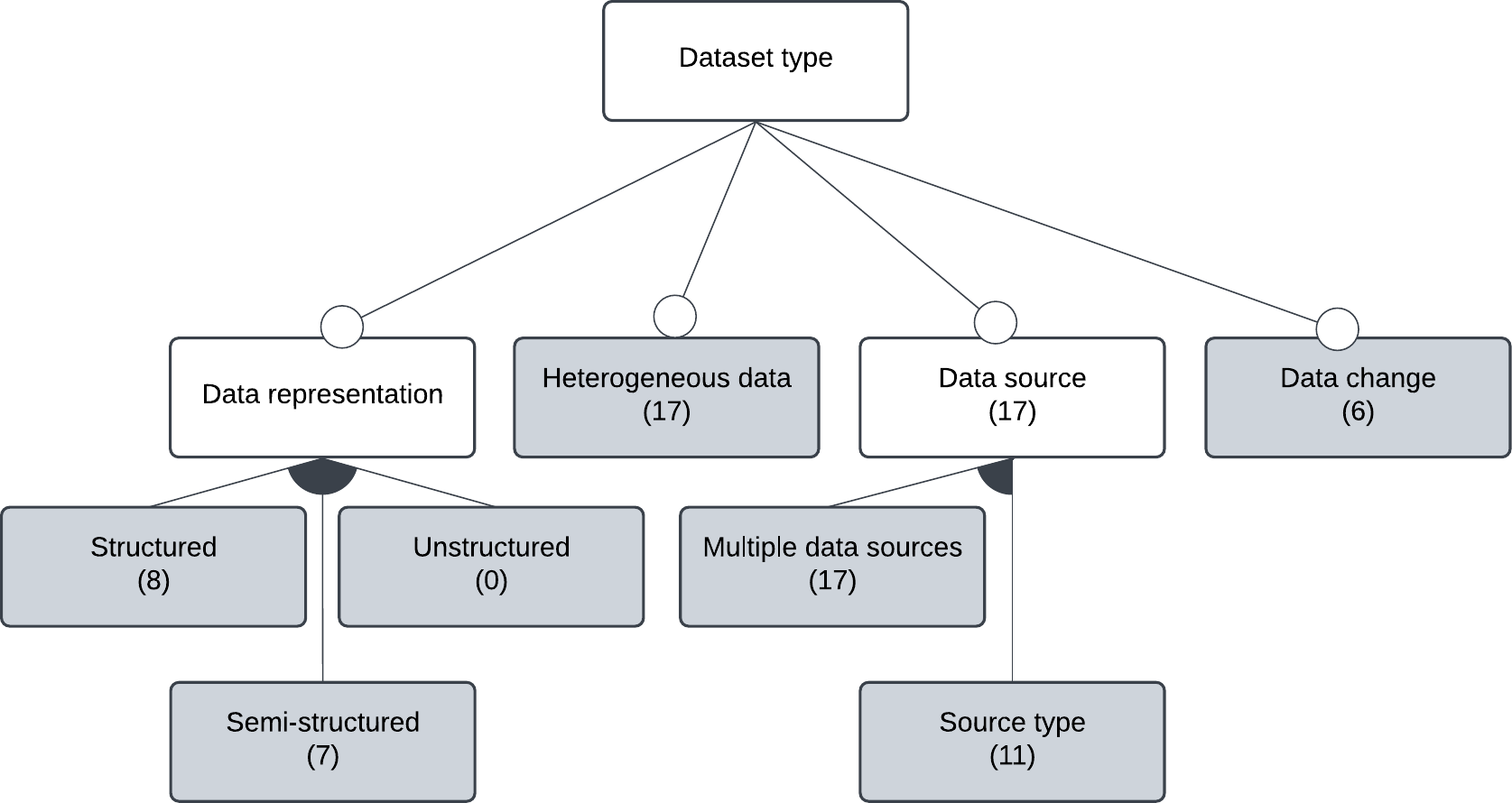}
  \caption{Dataset type feature of the FODA feature model for DQ definitions (legend: cf.~\autoref{fig:foda_legend}).}
  \label{fig:foda_model_dataset_type}
  \Description{The "dataset type" feature of the FODA feature model presented in this paper.}
\end{figure}

\begin{figure}[h]
  \centering
  \includegraphics[width=0.7\linewidth]{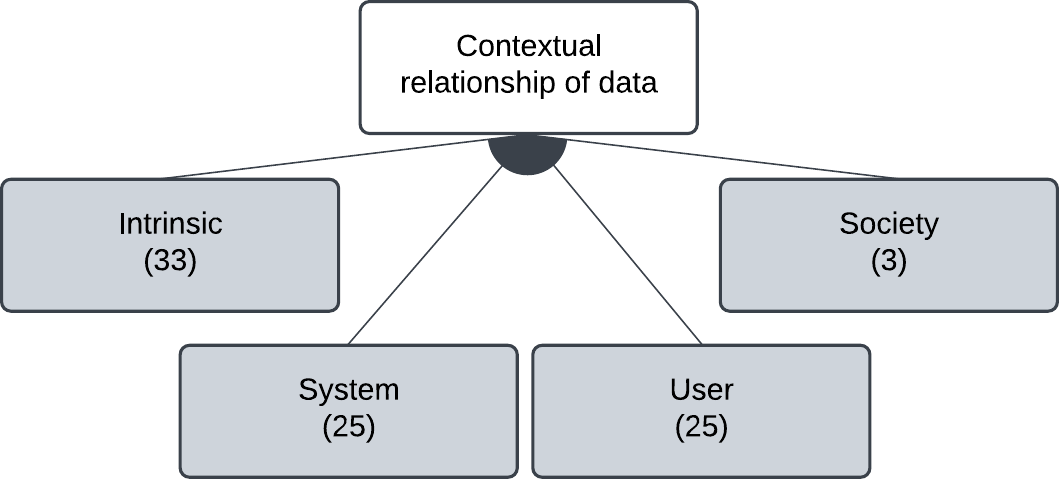}
  \caption{Contextual relationship feature of the FODA feature model for DQ definitions (legend: cf.~\autoref{fig:foda_legend}).}
  \label{fig:foda_model_contextual_relations}
  \Description{The feature "contextual relationship" feature of the FODA feature model presented in this paper.}
\end{figure}

\begin{figure}[h]
  \centering
  \includegraphics[width=\linewidth]{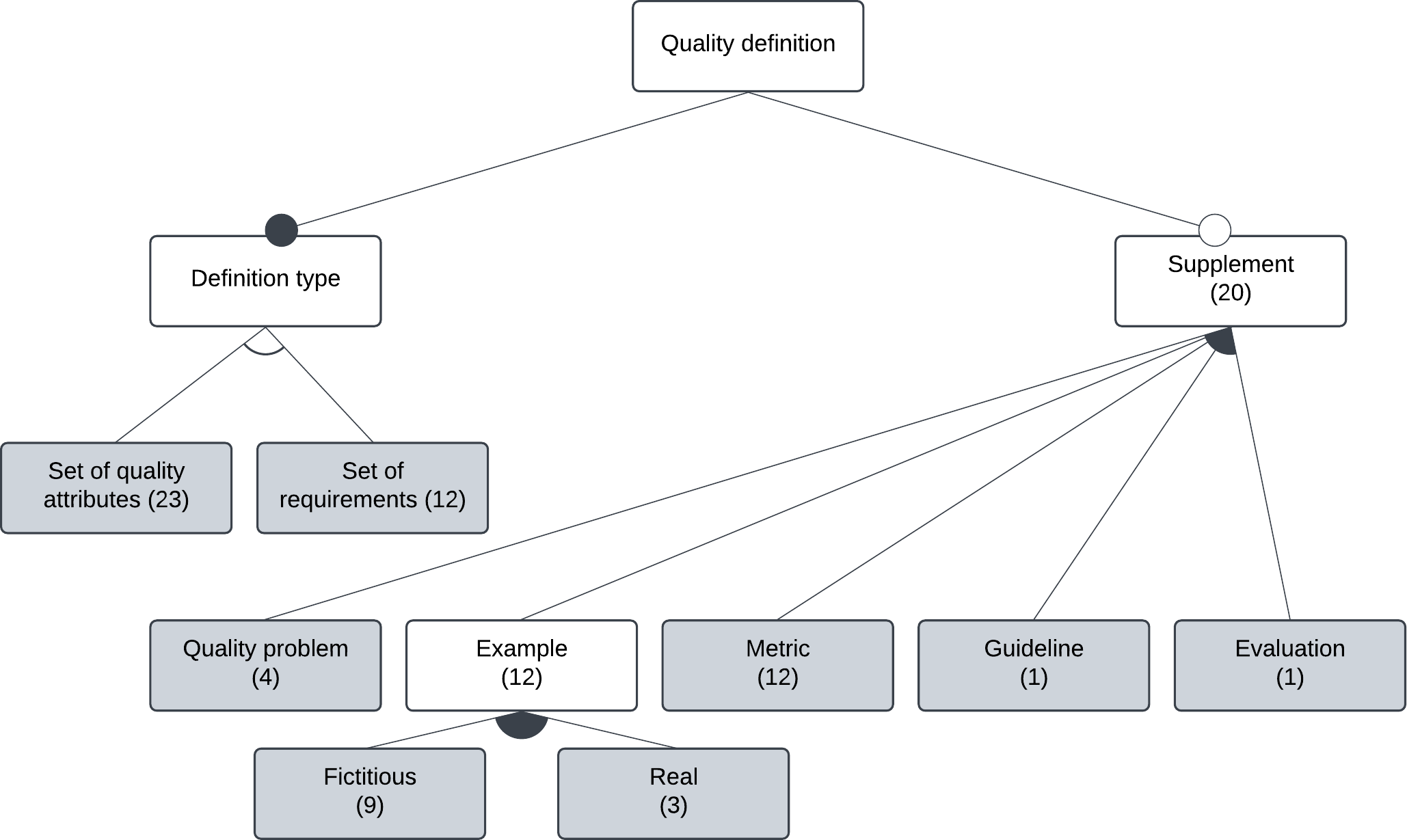}
  \caption{Quality definition feature of the FODA feature model for DQ definitions (legend: cf.~\autoref{fig:foda_legend}).}
  \label{fig:foda_model_quality_definition}
  \Description{The feature "quality definition" feature of the FODA feature model presented in this paper.}
\end{figure}

\begin{figure}[h]
  \centering
  \includegraphics[width=\linewidth]{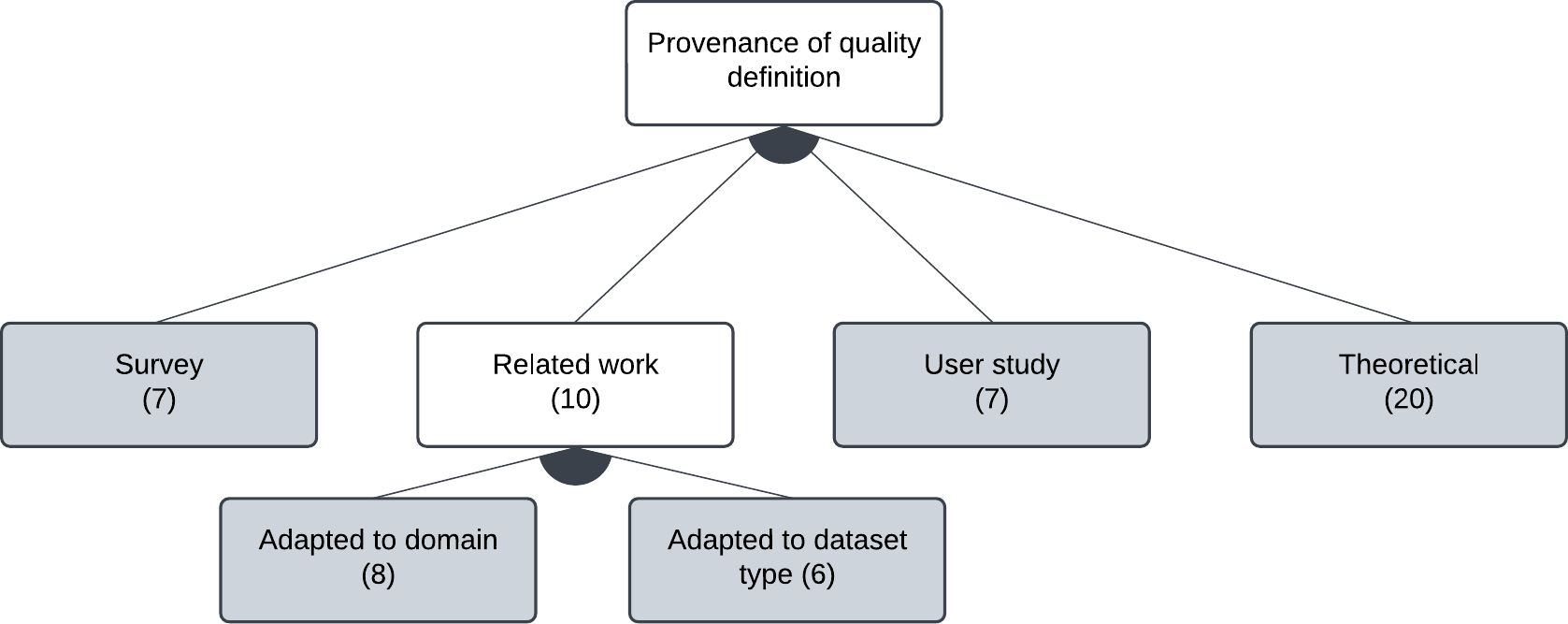}
  \caption{Provenance of quality definition feature of the FODA feature model for DQ definitions (legend: cf.~\autoref{fig:foda_legend}).}
  \label{fig:foda_model_provenance}
  \Description{The feature "provenance of quality definition" feature of the FODA feature model presented in this paper.}
\end{figure}

The feature model is presented in several diagrams (see \autoref{fig:foda_model_root} to \autoref{fig:foda_model_provenance}) and a detailed description below.
Each feature description includes a definition of the feature itself and any subfeatures, as well as a list of sample publications to which the feature applies.
Like Czarnecki and Helsen~\cite{czarnecki2006}, we do not intend the resulting terminology to be normative.
A full classification of all the publications in \autoref{tab:slr_publications} along the feature model is presented in \autoref{tab:foda_results}.
For each group of subfeatures, the feature descriptions below justify whether it forms an xor group or an or group.
An xor group denotes mutually exclusive subfeatures, whereas an or group denotes subfeatures that a publication may have in combination.
Both group types are collectively exhaustive for the publications that have the parent feature, since each of them has at least one subfeature.
If an optional parent feature is absent, the DQ definition does not address this aspect, which is shown as an empty cell in \autoref{tab:foda_results}.
Optional features outside of a group can be selected independently of each other, and each of them is either present or absent in a publication.
All three authors of this paper first performed the classification of the publications in \autoref{tab:slr_publications} independently and then discussed it to reach a consensus.

\paragraph{Data quality.} Since our feature model classifies DQ definitions, its root feature is called \emph{data quality}.
It has four \emph{mandatory} child features, which are designed along the research questions (RQ2.1) - (RQ2.4) and denote the aspects by which a DQ definition is characterized: the \emph{data} whose quality is being defined, the \emph{contextual relationship of data} that the definition assumes, the \emph{quality definition} itself, i.e., the way its quality dimensions are stated, and the \emph{provenance of quality definition}, i.e., the way these dimensions were derived.
Since every DQ definition is characterized by all four features, no number (n) is attached to them in \autoref{fig:foda_model_root}.
The four features are reference features shown in \autoref{fig:foda_model_root} and are refined in the following figures.

\paragraph{Data.}
Data is a multifaceted term with widely varying ideas and concepts.
The ISO 25012 standard~\cite{iso2008} provides a highly general definition of data as a ``re-interpretable representation of information in a formalized form suitable for communication, interpretation, or processing``~\cite{iso2008}.
Information is further defined as ``knowledge about objects, such as facts, events, things, processes, or ideas, including concepts, that have a particular meaning in a given context``~\cite{iso2008}.
Since we have to decide what and how to represent information in data, data provides us with a model of the world we are interested in.

Data quality definitions found in the literature vary widely in terms of the data for which quality is defined. While there are many quality definitions that do not further classify the data under consideration, others specify requirements for the \emph{dataset type}, the \emph{data model} if used, the \emph{domain of interest}, and the \emph{context of use}. As all child features of \emph{data} are optional, a DQ definition may select \emph{data} without any of them, as in Wang \& Strong~\cite{wang1996}, whose definition applies to data in general, whereas Zaveri et al.~\cite{zaveri2015} restrict all four of them. The feature model for \emph{data} is shown in \autoref{fig:foda_model_data} and explained below.

\paragraph{Dataset type.}
Data quality definitions often make certain assumptions about the data for which quality is defined.
The category of data that is considered is critical for an appropriate definition of DQ.
This feature is referred to as the \emph{dataset type}. Its refining feature model is shown in \autoref{fig:foda_model_dataset_type}.
We describe this feature with four child features: \emph{data representation}, \emph{heterogeneous data}, \emph{data source}, and \emph{data change}.
Early definitions of DQ such as Wand \& Wang~\cite{wand1996}, Wang \& Strong~\cite{wang1996}, Strong et al.~\cite{strong1997}, and Liu \& Chi~\cite{liu2002} did not specify the dataset type for which they defined quality.
Therefore, the feature \emph{dataset type} is optional, as are its child features, so that a DQ definition may address any combination of \emph{data representation}, \emph{heterogeneous data}, \emph{data source}, and \emph{data change}, or none of them at all.
Later, the type of dataset for which DQ is defined is explicitly considered more often. 

\paragraph{Data representation.}
Several classifications of data have been considered to define DQ.
When considering the \emph{representation of data}, as in the quality definition of Batini et al.~\cite{batini2009}, a distinction is made between \emph{structured data}, as in Montori et al \cite{montori2013} and Kulikowski \cite{kulikowski2014}, where data is represented in a precise structure that conforms to a data model, such as relational data, and \emph{semi-structured data}, which is represented using a data description language, such as XML, that pre-structures the data, as in Kandari et al. \cite{kandari2011} and Zaveri et al. \cite{zaveri2015}.
Semi-structured data may also have a data model, such as an XML schema, that further describes the data.
When XML data does not refer to an XML schema, there is no external representation of its structure, so it is said to be self-describing.
\emph{Unstructured data}, such as text, images, and video, exists without any explicit representation of its structure, either internally or externally.
The three subfeatures form an or group, since a DQ definition may address several data representations.
Batini et al.~\cite{batini2009}, for example, explicitly focus on both structured and semi-structured data, and Scannapieco et al.~\cite{scannapieco2004} define quality dimensions over conceptual schema elements, while the data are exchanged as XML documents.
None of the classified publications defines DQ for unstructured data, even where such data is mentioned as part of heterogeneous data, as in Kahn et al.~\cite{kahn2016}.
We nevertheless retain the feature \emph{unstructured} to complete the established distinction of data representations, whose absence reflects the focus of DQ research on structured and semi-structured data observed by Batini et al.~\cite{batini2009}.

\paragraph{Heterogeneous data.}
\emph{Heterogeneous data}
is any data with a high variability of data representations.
This applies, for example, to Jarke et al.~\cite{jarke1999}, Scannapieco et al.~\cite{scannapieco2004}, Klein et al.~\cite{klein2007streaming}, Batini et al.~\cite{batini2009}, and Zaveri et al.~\cite{zaveri2015}.
Due to the diversity of data sources, such as in the Internet of Things, e.g., Cai \& Zhu~\cite{cai2015}, the data collected are often of different types with heterogeneity.
For example, electronic health records combine structured patient data with unstructured findings from medical examinations (Kahn et al.~\cite{kahn2016}).
Heterogeneity is nevertheless distinct from the features \emph{data representation} and \emph{data source}.
It may occur within a single representation type, as in Verma~\cite{verma2006} and Klein et al.~\cite{klein2007streaming}, and without multiple data sources, as in Scannapieco et al.~\cite{scannapieco2004} and DeCastro-García \& Pinto~\cite{decastro2021}.
Conversely, Ballou \& Pazer~\cite{ballou1985} and Montori et al.~\cite{montori2013} consider multiple data sources without heterogeneous data (see \autoref{tab:foda_results}).
\paragraph{Data source.}
The \emph{data source}
refers to the origin or location from which data is collected, retrieved or stored.
A \emph{source type} refers to the classification or category of the data source based on its characteristics, format, or method of access.
Data from \emph{multiple data sources} can be of different \emph{source types}, e.g., health data typically includes structured patient information data as well as unstructured data from medical examinations (e.g., Kahn et al.~\cite{kahn2016}).
Since multiple data sources are typically not synchronized, they can be ambiguous and contain data redundancies.
The source types addressed in the classified publications include sensors and connected devices, e.g., Cai \& Zhu~\cite{cai2015} and Montori et al.~\cite{montori2013}, data warehouses integrating operational databases, e.g., Jarke et al.~\cite{jarke1999}, web sources, e.g., Kandari et al.~\cite{kandari2011}, and clinical information systems, e.g., Kahn et al.~\cite{kahn2016}.
The subfeatures \emph{multiple data sources} and \emph{source type} form an or group, since a DQ definition that addresses the data source considers multiple sources, the type of source, or both.
In the classified publications, source types are only addressed in combination with multiple data sources, but a definition may equally address the type of a single source, such as a single sensor.

\paragraph{Data change.}
Data can also be classified according to \emph{data change}.
Data that is not dynamic is considered either static (unchanging) or persistent, which is data that is accessed infrequently and is unlikely to change.
A released official-statistics dataset that remains unchanged after publication is an example of static data, whereas continuously updated social media data, as analyzed by El Alaoui et al.~\cite{elalaoui2019}, is dynamic.
While static and dynamic data is stored before it is processed, streaming data is a constant flow of data that is processed before it is stored.
In an increasingly connected world, data is often distributed across multiple data sources that store and link heterogeneous data types, including static data, dynamic data, and streaming data.
Klein et al.~\cite{klein2007streaming} used the quality definition for static data to define the quality of streaming data, and Klein~\cite{klein2007sensor} considered quality aspects of sensor data streams.
Data change is also a subject in Ballou \& Pazer~\cite{ballou1985}, Cai \& Zhu~\cite{cai2015}, Ceravolo \& Bellini~\cite{ceravolo2019}, and El Alaoui et al.~\cite{elalaoui2019}.
\paragraph{Data model.}
Structured data conforms to a \emph{data model} that explicitly defines its structure.
While structured data always has a data model, semi-structured data may or may not have a data model.
A data model is a graphical and/or lexical representation model that specifies the properties, structure, and inter-relationships of data in a domain of interest.
In the context of relational or XML data, a data model is also called a schema.
We assign the feature \emph{data model} only if a DQ definition explicitly refers to the data model of the data.
Hence, a definition for structured data does not necessarily have this feature, as in Ballou \& Pazer~\cite{ballou1985} and Klein~\cite{klein2007sensor}.
Data models play a role in the DQ definitions of Jarke et al.~\cite{jarke1999}, Scannapieco et al.~\cite{scannapieco2004}, Verma~\cite{verma2006}, Batini et al.~\cite{batini2009}, Ge et al.~\cite{ge2011}, Chiasera et al.~\cite{chiasera2011}, Kulikowski~\cite{kulikowski2014}, Zaveri et al.~\cite{zaveri2015}, Cai \& Zhu~\cite{cai2015}, and Kahn et al.~\cite{kahn2016}.
The data models these definitions refer to differ considerably, ranging from the data warehouse schemas of Jarke et al.~\cite{jarke1999} to the RDF vocabularies underlying the linked data quality definition of Zaveri et al.~\cite{zaveri2015}.
DQ considerations can also be extended to the quality of data models.
For example, Moody \& Shanks~\cite{moody1994makes} evaluated the quality of schemas defined as entity-relationship models.

\paragraph{Domain of interest.}
In general, data represents information about a selected \emph{domain of interest}.
Such a domain may be a segment of the real world or a virtual domain, such as digitized objects.
Data quality has also been defined for several specific domains of interest.
For example, for healthcare, e.g., Kahn et al.~\cite{kahn2016}, Chiasera et al.~\cite{chiasera2011}, the Internet of Things, e.g., Montori et al.~\cite{montori2013}, cybersecurity, e.g., DeCastro-García \& Pinto~\cite{decastro2021}, scientific publications, e.g., Azeroual \& Lewoniewski~\cite{azeroual2020}, the World Wide Web, e.g., Kandari et al.~\cite{kandari2011}, Caro et al.~\cite{caro2007}, low-carbon information quality, e.g., Gao et al.~\cite{gao2024}, statistical agencies, e.g., Brackstone et al.~\cite{brackstone1999}, and crisis management, e.g., Friberg et al.~\cite{friberg2011}.
Data quality is also considered independent of any domain in various publications, such as Ballou \& Pazer~\cite{ballou1985}, Wang \& Strong~\cite{wang1996}, Batini et al.~\cite{batini2009}.
In the feature model, we distinguish only between data that are \emph{specific} and \emph{independent} of a particular \emph{domain of interest}, leaving the feature model open to other domains.

\paragraph{Context of use.}
Data selection and presentation may depend on the \emph{context of use}, which includes the questions to be answered, the tasks to be performed to answer those questions, the processing of the data, and the tools to be used to perform those tasks with the data.
To make data \emph{fit for use}, each context of use needs to be examined for specific DQ requirements.
Typical examples of data processing tasks that provide a specific context for DQ considerations include data search and analysis, such as Azeroual \& Lewoniewski~\cite{azeroual2020}, El Alaoui et al.~\cite{elalaoui2019}, Zaveri et al.~\cite{zaveri2015}, Kandari et al.~\cite{kandari2011}, clinical auditing, e.g., Verma~\cite{verma2006}, data mining, e.g., Reddy et al.~\cite{reddy2007}, decision-making in crisis management, e.g., Friberg et al.~\cite{friberg2011}, and digital marketing, e.g., Gao et al.~\cite{gao2024}.
When considering DQ, the context of use is also often not specified, as in Wang \& Strong~\cite{wang1996}, Jarke et al.~\cite{jarke1999}, and Batini et al.~\cite{batini2009}, to name a few.
In the feature model, we only distinguish between data that is \emph{specific} and \emph{independent} of a particular \emph{context of use}, leaving the feature model open to other contexts of use.

\paragraph{Contextual relationship of data.}
When DQ refers only to the data itself, it is called \emph{intrinsic}. Because data do not stand alone, we consider their relationships with context, such as the data \emph{users}, the \emph{system} in which the data are used, and last but not least, the \emph{society} (or community) in which the data were collected.
The corresponding part of the feature model is shown in \autoref{fig:foda_model_contextual_relations}.

Some publications classify all the quality dimensions they consider into a few groups, such as Wang \& Strong~\cite{wang1996}, Wand \& Wang~\cite{wand1996}, ISO 25012~\cite{iso2008}, Zaveri et al.~\cite{zaveri2015}, Ceravolo \& Bellini~\cite{ceravolo2019}.
These groups will be discussed in more detail below.

When DQ refers only to the data itself, it is called \emph{intrinsic}, as defined by e.g. Wang \& Strong~\cite{wang1996} and Zaveri et al.~\cite{zaveri2015}.
This type of quality is also called \emph{internal} or data-related, as by Wand \& Wang~\cite{wand1996}.
In ISO 25012~\cite{iso2008} it is called \emph{inherent}, but includes considerably more quality dimensions, in fact some dimensions can also be considered as related to the system or the user.
Ceravolo \& Bellini~\cite{ceravolo2019} define a syntactic level of quality that ``can be validated by internal observations``~\cite{ceravolo2019}.
A typical dimension of intrinsic quality is accuracy (see Wang \& Strong~\cite{wang1996} and Zaveri et al.~\cite{zaveri2015}).
However, Ceravolo \& Bellini~\cite{ceravolo2019} map accuracy to a semantic level of quality that ``can be validated by external observations``~\cite{ceravolo2019}.
Consistency is also mentioned as an intrinsic quality dimension in Wand \& Wang~\cite{wand1996} and Zaveri et al.~\cite{zaveri2015}.
Completeness is a dimension that is classified differently in the literature.
While it is considered as intrinsic in Zaveri et al.~\cite{zaveri2015} (or inherent in ISO 25012~\cite{iso2008}), Wang \& Strong~\cite{wang1996} classify it as contextual in the sense that completeness has to be considered in the context of use, so it can also have a system or user relationship.
We classified the publications according to the contextual relationships of  the completeness dimension found in their specific DQ definitions.

Strong et al.~\cite{strong1997} suggested that data are of high quality when they meet the expectations of the users, which can be humans or systems, and define this data as \emph{fit for use by data consumers}.
The user-related category of DQ includes the ease with which the data can be understood without ambiguity in a given context of use, and the extent to which the data are available in different forms for different cultural perspectives, different technologies for accessing the data, and for different sensory abilities of the user.
The credibility or trustworthiness of the data is a further important aspect of the relationship between data and users.
Wand \& Wang~\cite{wand1996} define an external view of data to consider dimensions such as relevance, conciseness, usability, and understandability.
The ISO 25012 standard (2008) classifies credibility and understandability as intrinsic, partly with system dependency.
Zaveri et al.~\cite{zaveri2015} consider relevance, trustworthiness, and understandability as contextual dimensions that have a user relationship.
Ceravolo \& Bellini~\cite{ceravolo2019} define a pragmatic level of DQ that ``refers to the assessment of properties related to fitness to intended uses``~\cite{ceravolo2019}, which includes, for example, relevance and usability.\\*
Quality can also relate to the \emph{system} that provides data, as data and data changes need to be available in the right context of use and with appropriate access rights for users.
For example, availability and security of data play an important role (mentioned in the accessibility group in Zaveri et al.~\cite{zaveri2015}).
If there is data from multiple sources, the system must integrate these sources so that the user can make use of all of them.
Zaveri et al.~\cite{zaveri2015} also consider interlinking in the accessibility group.
As data may change over time, there is also a need to keep data up-to-date and traceable over time.
The ISO 25012 standard~\cite{iso2008} mentions traceability as a system-dependent (and also inherent) quality dimension.
Zaveri et al.~\cite{zaveri2015} mention timeliness as a contextual dimension related to the system, but do not consider traceability at all.
\\*
Last but not least, DQ also relates to the \emph{society} in which the data were collected, which leads to ethical questions.
This relationship is considered by Shanks \& Corbitt~\cite{shanks1999}, Montori et al.~\cite{montori2013}, and Firmani et al.~\cite{firmani2020}.
When considering a dataset, quality aspects such as lack of bias, data provenance and diversity relate to data selection as well as to data integration and knowledge extraction.
Wand \& Wang~\cite{wand1996} already mentioned freedom from bias as part of their external view of data.

We assigned a publication to a specific contextual relationship if its DQ definition directly mentioned this contextual relationship or if we interpreted the definition accordingly.
The four subfeatures form an or group.
Every DQ definition has at least one of these relationships, and most definitions consider several of them, as \num{28} of the \num{35} publications do (see \autoref{tab:foda_results}).
Restricting the classification to a single contextual relationship would therefore misrepresent these definitions.

\paragraph{Quality definition.}
There is also considerable variation in how DQ is defined.
A quality framework such as Scannapieco et al.~\cite{scannapieco2004} and Zaveri et al.~\cite{zaveri2015} typically includes a declarative description of DQ along with quality metrics.
This is very much in line with the goal-question-metric approach of Basili et al.~\cite{basili1994}, where a quality dimension is first defined declaratively.
A list of quality attributes (belonging to this dimension) can then be used to formulate quantifiable questions about DQ that metrics are intended to answer.
Although quality metrics are useful for measuring certain aspects of quality, the data engineer must always be aware that DQ remains a qualitative concept that cannot be fully measured.
Therefore, we do not consider a pure set of metrics to be the definition of DQ, but rather expect a declarative formulation of DQ requirements or attributes.
The feature model for \emph{quality definition} is shown in \autoref{fig:foda_model_quality_definition}.
In the following, we distinguish the \emph{type of quality definition} and the \emph{supplement} as subfeatures.

\paragraph{Definition type.}
Ideally, DQ is defined by formulating a \emph{set of requirements} that the data must meet.
A \emph{requirement} for a quality dimension is defined in a declarative way with a goal to be achieved.
For example, completeness refers to the degree to which a dataset contains all data elements of the domain of interest.
In many publications, we instead find a \emph{set of quality attributes} that describes the quality of the data.
These attributes are often accompanied by only short descriptions.
An attribute-based definition describes a dimension by listing adjectives.
For example, completeness is then described as sufficient and comprehensive, without stating the condition that the data must meet.
Quality attributes are presented with or without explanations in, e.g., Wang \& Strong~\cite{wang1996}, Strong et al.~\cite{strong1997}, Liu \& Chi~\cite{liu2002}, Kandari et al.~\cite{kandari2011}, and Kahn et al.~\cite{kahn2016}.
Quality requirements are presented in, e.g., Ballou \& Pazer~\cite{ballou1985}, Wand \& Wang~\cite{wand1996}, Scannapieco et al.~\cite{scannapieco2004}, ISO 25012~\cite{iso2008}, and Zaveri et al.~\cite{zaveri2015}.\\*
Ballou \& Pazer~\cite{ballou1985} and Wand \& Wang~\cite{wand1996} define multiple quality dimensions through mappings between the data or information system and a real-world system, which also leads to to a quality definition by requirements.
In summary, it can be said that requirements specify the quality goal to be achieved, whereas quality attributes are only mentioned or shortly described.
\emph{Definition type} is mandatory, since we require every DQ definition to be formulated declaratively, either as requirements or as attributes.
The two subfeatures form an xor group, since a DQ definition either specifies the goals that its quality dimensions must achieve or only names and describes them.

\paragraph{Supplement.}
DQ definitions are typically also supported by supplementary evidence.
\emph{Supplement} is optional, since such evidence supports a definition but is not part of it, and \num{15} of the \num{35} publications provide none (see \autoref{tab:foda_results}).
We distinguish between \emph{example}, \emph{metric}, \emph{quality problem}, \emph{guideline}, and \emph{evaluation}.
These subfeatures form an or group, since a publication may support its definition with several kinds of supplements, e.g., Zaveri et al.~\cite{zaveri2015} provide quality problems, real examples, and metrics.\\*
To illustrate aspects of DQ, an \emph{example} is often given that is \emph{fictitious} or \emph{real}.
The two subfeatures also form an or group, since nothing prevents a publication from combining both kinds of examples, even though none of the classified publications does so.
Fictitious examples are artificially created for illustrative purposes only, as in Strong et al.~\cite{strong1997}, Liu \& Chi~\cite{liu2002}, Klein~\cite{klein2007sensor}, Klein et al.~\cite{klein2007streaming}, ISO 25012~\cite{iso2008}, DAMA UK Working Group~\cite{dama2013}, Montori et al.~\cite{montori2013}, Azeroual \& Lewoniewski~\cite{azeroual2020}, and DeCastro-García \& Pinto~\cite{decastro2021}.
Real examples are derived from authentic, real-world datasets and can demonstrate relevance in practice, as in Shanks \& Corbitt~\cite{shanks1999}, Scannapieco et al.~\cite{scannapieco2004}, and Zaveri et al.~\cite{zaveri2015}.\\*
It may also help to illustrate DQ by listing metrics, quality problems and guidelines.
Essentially, a \emph{metric} is a function used to measure DQ.
Such a metric is typically based on a numerical definition.
For example, completeness can be measured as the ratio of the data elements actually present to all elements expected for the domain of interest (Zaveri et al.~\cite{zaveri2015}).
Publications that both define dimensions and provide metrics include Jarke et al.~\cite{jarke1999}, Scannapieco et al.~\cite{scannapieco2004}, ISO 25012~\cite{iso2008}, Batini et al.~\cite{batini2009}, DAMA UK Working Group~\cite{dama2013}, and Zaveri et al.~\cite{zaveri2015}.
\emph{Quality problems} occur when data does not meet the defined quality (i.e., Wand \& Wang~\cite{wand1996}, Strong et al.~\cite{strong1997}, Caro et al.~\cite{caro2007}, and Zaveri et al.~\cite{zaveri2015}).
A \emph{guideline} is a directive or instruction designed to prevent or reduce quality problems.
An example for guidelines for improving statistical data can be found in Brackstone~\cite{brackstone1999}.
In addition, guidelines play a central role in two closely related publications: the FAIR principles by Wilkinson et al.~\cite{wilkinson2016} and CARE principles by Caroll et al.~\cite{carroll2020}.\\*
An \emph{evaluation} of a dimension assesses the individual dimension based on the definition.
There is an \emph{evaluation} of a quality definition when it is applied in an empirical study.
This was done, for example, in Strong et al.~\cite{strong1997}.

\paragraph{Provenance of quality definition.}
For all publications, we examine how the quality definition was developed.
In particular, we are interested in the information from which the definition is derived.
The corresponding feature model is shown in \autoref{fig:foda_model_provenance}.
Its subfeatures form an or group, since every DQ definition is derived in at least one way and several ways are often combined.
For example, Batini et al.~\cite{batini2009} build on a survey, adapt existing definitions to specific dataset types, and develop their definition theoretically.

Often a publication starts with a review of the \emph{related work}, i.e. publications that already present DQ definitions.
Some publications present a systematic literature survey that compares publications along a classification scheme, such as Zaveri et al.~\cite{zaveri2015}, Ge et al.~\cite{ge2011}, Kandari et al.~\cite{kandari2011}, and Schaal et al.~\cite{schaal2012}.
This should help to identify parts of quality definitions that can be reused in a new definition.
Zaveri et al.~\cite{zaveri2015} show very well how to use a conducted systematic literature survey as a prerequisite for their own definition of DQ.
When quality definitions are presented for \emph{specific domains} or \emph{specific dataset types}, such a definition is sometimes derived from a more general DQ definition that is cited as related work.
Caro et al.~\cite{caro2007}, Reddy et al.~\cite{reddy2007}, Chiasera et al.~\cite{chiasera2011}, Kandari et al.~\cite{kandari2011}, Kahn et al.~\cite{kahn2016}, El Alaoui et al.~\cite{elalaoui2019}, Azeroual \& Lewoniewski~\cite{azeroual2020}, and DeCastro-García \& Pinto~\cite{decastro2021} \emph{adapt} general DQ definitions to more \emph{specific domains} such as healthcare, the Internet of Things, scientific publications, and cybersecurity.\\*
Jarke et al.~\cite{jarke1999}, Reddy et al.~\cite{reddy2007}, Batini et al.~\cite{batini2009}, Chiasera et al.~\cite{chiasera2011}, Kandari et al.~\cite{kandari2011}, and Kahn et al.~\cite{kahn2016} \emph{adapted} general DQ definitions to \emph{specific dataset types} such as data warehouses and data streams.
The two kinds of adaptation form an or group, since a definition may be adapted to a specific domain and a specific dataset type at the same time, as in Chiasera et al.~\cite{chiasera2011} and Kahn et al.~\cite{kahn2016}.
For example, the quality dimensions in Wang \& Strong~\cite{wang1996} were used as a basis for adaptation in Kandari et al.~\cite{kandari2011} and Schaal et al.~\cite{schaal2012}.
The quality dimensions in Cai et al.~\cite{cai2015} were used in El Alaoui et al.~\cite{elalaoui2019}.\\*
The proper development of the DQ definition can be done empirically by conducting a \emph{user study} or \emph{theoretically} by developing a declarative quality definition based on fundamental concepts.
In user studies, data consumers are asked to identify the characteristics they use to assess whether data are fit for use in their tasks (see e.g., Wang \& Strong~\cite{wang1996}, Strong et al.~\cite{strong1997}, Caro et al.~\cite{caro2007}, Friberg et al.~\cite{friberg2011}, and Pansara~\cite{pansara2023}).
Examples of theoretical definitions include Ballou \& Pazer~\cite{ballou1985}, Brackstone~\cite{brackstone1999}, Liu \& Chi~\cite{liu2002}, Scannapieco et al.~\cite{scannapieco2004}, Klein~\cite{klein2007sensor}, Batini et al.~\cite{batini2009}, ISO 25012~\cite{iso2008}, Cai \& Zhu~\cite{cai2015}, and Zaveri et al.~\cite{zaveri2015}.
A theoretical definition of DQ includes the definition of fundamental concepts, especially of the type of data considered. The definition of the DQ dimensions is then based on these defined concepts. 
The degree to which the theoretical definition is formally given is not addressed by this classification, as our focus is on the provenance of a DQ definition.

\subsection{Classification Results}
\label{sec:foda_classification}

Given the feature model described in \autoref{sec:foda_feature_model}, the publications listed in \autoref{tab:slr_publications} are to be classified according to this feature model.
Next, we present our classification and highlight the specialties of this classification.

\subsubsection{Classification of selected publications.}
\label{sec:foda_legend}
 We classified all publications listed in \autoref{tab:slr_publications} according to our feature model described in \autoref{sec:foda_feature_model}.
 \autoref{tab:foda_results} shows the result of this feature-oriented domain analysis.
The following legend explains the entries in \autoref{tab:foda_results}. The columns of this table (except the leftmost column) represent essential features. Each entry either represents the selected subfeatures or indicates whether the feature itself is selected (indicated by an \emph{x}).
For all columns, an \emph{empty cell} means that the corresponding feature is \emph{undefined / not selected}, while multiple entries indicate multiple selected subfeatures, separated by a plus sign (+). For each column, its name and the possible entries are listed below. 
\\*

\noindent \emph{Representation}: s = Structured; m = Semi-structured; u = Unstructured\\*
\emph{Heterogeneous}: x = The data addressed is heterogeneous\\*
\emph{Data source}: m = Multiple sources are addressed; t = Type of data source is addressed\\*
\emph{Data change}: x = Data change is addressed\\*
\emph{Data model}: x = The data is structured by a data model\\*
\emph{Domain of interest}: x = Specific domain is addressed\\*
\emph{Context of use}: x = Specific context of use is addressed\\*
\emph{Contextual relationship of data}: i = Intrinsic; s = System; u = User; o = Society\\*
\emph{Definition type}: a = Set of attributes; r = Set of requirements\\*
\emph{Supplement}: p = Quality problem; f = Fictitious example; r = Real example; m = Metric; g = Guideline; e = Evaluation\\*
\emph{Provenance}: s = Survey; d = Related work adapted to domain ; y = Related work adapted to dataset type; u = User study; t = Theoretical

\begin{table*}
\caption{Classification results of publications in accordance with the feature model (legend: cf. \autoref{sec:foda_legend})}
\label{tab:foda_results}
% \footnotesize
\setlength{\tabcolsep}{3pt}
\begin{tabular}{@{}lccccccccccc@{}}
    \toprule
    & \multicolumn{7}{c}{\emph{Data}} & & \multicolumn{2}{c}{\emph{Quality definition}} & \\
    \cmidrule(lr){2-8} \cmidrule(lr){10-11}
    Authors \& Publication & \begin{turn}{90}Representation\end{turn} & \begin{turn}{90}Heterogeneous\end{turn} & \begin{turn}{90}Data source\end{turn} & \begin{turn}{90}Data change\end{turn} & \begin{turn}{90}Data model\end{turn} & \begin{turn}{90}Domain of\end{turn}\begin{turn}{90}interest\end{turn} & \begin{turn}{90}Context of use\end{turn} & \begin{turn}{90}Contextual\end{turn}\begin{turn}{90}relationship\end{turn} & \begin{turn}{90}Definition type\end{turn} & \begin{turn}{90}Supplement\end{turn} & \begin{turn}{90}Provenance\end{turn} \\
    \midrule
    Ballou \& Pazer~\cite{ballou1985}               & s     &   & m     & x &   &   &   & i+s   & r     &       & t     \\
    Wang \& Strong~\cite{wang1996}                  &       &   &       &   &   &   &   & i+s+u & a     &       & u     \\
    Wand \& Wang~\cite{wand1996}                    &       &   &       &   &   &   &   & i+s+u & r     & p     & t     \\
    Strong et al.~\cite{strong1997}                 &       &   &       &   &   &   &   & i+s+u & a     & p+f+e & u     \\
    Jarke et al.~\cite{jarke1999}                   &       & x & m     &   & x &   &   & i+s+u & a     & m     & y     \\
    Shanks \& Corbitt~\cite{shanks1999}             &       &   &       &   &   &   &   & i+u+o & a     & r     & t     \\
    Brackstone~\cite{brackstone1999}                &       &   &       &   &   & x &   & i+s+u   & r     & g     & t     \\
    Liu \& Chi~\cite{liu2002}                       &       &   &       &   &   &   &   & i+s+u & a     & f     & t     \\
    Scannapieco et al.~\cite{scannapieco2004}       & s+m   & x &       &   & x &   &   & i     & r     & r+m   & t     \\
    Verma~\cite{verma2006}                          & s     & x & m+t   &   & x & x & x & i+s   & a     &       & t     \\
    Caro et al.~\cite{caro2007}                     & m     & x & m+t   &   &   & x & x & i+s+u & a     & p+m   & d+u   \\
    Reddy et al.~\cite{reddy2007}                   & m     & x & m+t   &   &   & x & x & i+s+u & a     &       & d+y   \\
    Klein et al.~\cite{klein2007streaming}          & s     & x & m+t   & x &   &   &   & i     & r     & f     & t     \\
    Klein~\cite{klein2007sensor}                    & s     &   &       & x &   &   &   & i     & r     & f     & t     \\
    ISO 25012~\cite{iso2008}                        &       &   &       &   &   &   &   & i+s+u & r     & f+m   & t     \\
    Batini et al.~\cite{batini2009}                 & s+m   & x & m     &   & x &   &   & i     & r     & m     & s+y+t \\
    Kandari et al.~\cite{kandari2011}               & m     & x & m+t   &   &   & x & x & i+s+u & a     &       & s+d+y \\
    Ge et al.~\cite{ge2011}                         &       &   &       &   & x &   &   & i+s+u & a     &       & s     \\
    Chiasera et al.~\cite{chiasera2011}             & m     & x & m+t   &   & x & x & x & i+s+u & a     & m     & d+y   \\
    Friberg et al.~\cite{friberg2011}               &       &   &       &   &   & x & x & i+s+u & a     &       & u     \\
    Schaal et al.~\cite{schaal2012}                 &       &   &       &   &   & x &   & u     & a     &       & s     \\
    Montori et al.~\cite{montori2013}               & s     &   & m+t   &   &   & x & x & i+u+o & a     & f+m   & s     \\
    DAMA UK WG~\cite{dama2013}                      &       &   &       &   &   &   &   & i+s   & r     & f+m   & t     \\
    Kulikowski~\cite{kulikowski2014}                & s     & x & m+t   &   & x &   &   & i+s+u & a     &       & s+t   \\
    Zaveri et al.~\cite{zaveri2015}                 & m     & x & m+t   &   & x & x & x & i+s+u & r     & p+r+m & s+t   \\
    Cai \& Zhu~\cite{cai2015}                       &       & x & m+t   & x & x &   &   & i+s+u & a     &       & t     \\
    Kahn et al.~\cite{kahn2016}                     &       & x & m+t   &   & x & x & x & i     & a     &       & d+y+u \\
    Ceravolo \& Bellini~\cite{ceravolo2019}         &       & x & m     & x &   &   &   & i+s+u & a     & m     & t     \\
    El Alaoui et al.~\cite{elalaoui2019}            &       & x & m     & x &   & x & x & i+s+u & a     & m     & d     \\
    Firmani et al.~\cite{firmani2020}               &       &   &       &   &   &   &   & o     & r     &       & t     \\
    Azeroual \& Lewoniewski~\cite{azeroual2020}     &       & x & m     &   &   & x & x & i+u   & a     & f     & d     \\
    Black \& Nederpelt~\cite{black2020}             &       &   &       &   &   &   &   & i+s+u & r     &       & t     \\
    DeCastro-García \& Pinto~\cite{decastro2021}    &       & x &       &   &   & x &   & i+s+u & a     & f+m   & d+t   \\
    Pansara~\cite{pansara2023}                      &       &   &       &   &   &   &   & i+s   & a     &       & u     \\
    Gao et al.~\cite{gao2024}                       &       &   &       &   &   & x & x & i+s+u & a     &       & u+t   \\
    \bottomrule
\end{tabular}
\end{table*}

\subsubsection{Similarities and differences within feature groups}
Considering our classification results, we need to interpret them and indicate any potential research gaps. We also discuss combinations of features that have not yet been explored.

\begin{figure}[h]
  \centering
  \includegraphics[width=\linewidth]{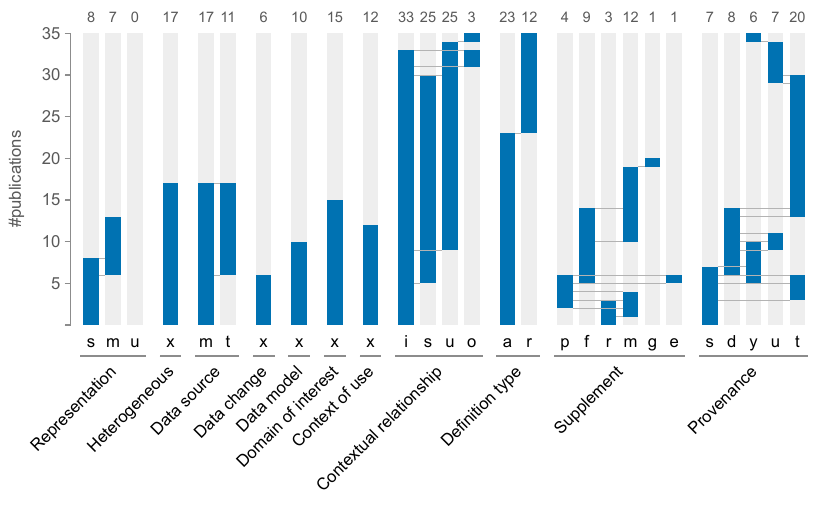}
  \caption{Classification of features: Frequency of features per publication, highlighting similarities and differences within feature groups. X-Axis: Features (abbreviations cf. \autoref{sec:foda_legend}). Y-Axis: Number of publications.}
  \label{fig:foda_result}
  \Description{This chart visualizes the frequency of features per publication, with all features on the X-Axis and the number of publications corresponding to each feature on the Y-Axis. The sum of all publications that match a specific feature is indicated at the top of the X-Axis. In contrast to a conventional bar chart, this chart highlights the similarities and differences between features.}
\end{figure}

As a preparation, \autoref{fig:foda_result} visualizes the frequency of publications per feature, with all features on the X-axis and the number of publications having this feature on the Y-axis.
The sum of all publications that have a specific feature is shown at the top of the X-axis.
The features are grouped into feature groups, e.g., the feature group \emph{representation} contains the features \emph{structured} (s), \emph{semi-structured} (m), and \emph{unstructured} (u).
Within each feature group, the occurrences are represented by blue bars. These bars do not necessarily start from the bottom, but are shifted to highlight the similarities and differences.
For example, looking at the feature group \emph{representation}, the overlap of the first two bars shows that there are \num{2} publications that deal with both structured and semi-structured data.
In other feature groups, the bars are partially interrupted to show overlap with some, but not necessarily all, of the other features in the group.
For example, the interruption of the feature \emph{society} (o) in the feature group \emph{contextual relationship of data} reveals \num{2} overlaps.
The upper part of the bar visualizes exactly \num{1} publication that addresses the feature \emph{society} (o) exclusively.
The lower part of the bar shows an overlap with the features \emph{intrinsic} (i) and \emph{user} (u), but not with the feature \emph{system} (s).

Looking at \autoref{tab:foda_results} and \autoref{fig:foda_result}, we would like to highlight the following specialties.
\begin{enumerate}
    \item Dataset type: \num{22} out of \num{35} publications do not mention any data representation feature, i.e. the quality definition applies to data in general.
    \item Data model: \num{25} publications do not mention the existence of a data model.
    \item Domain of interest: \num{15} publications define DQ for a specific domain of interest.
    \item Context of use: \num{12} publications mention a specific context of use.
    \item Contextual relationship: \num{21} publications consider the combination of intrinsic, system, and user as contextual relationships, only \num{3} publications mention society as a contextual relationship. \num{5} publications focus on intrinsic DQ only.
    \item Definition type: Only \num{12} publications define DQ in terms of requirements.
    \item Supplement: Only \num{3} publications include real examples, \num{9} publications include fictitious examples, \num{23} publications do not include any examples. \num{12} publications include quality metrics. (Note that we have not examined whether subsequent papers include more supplements.)
    \item Provenance: \num{7} publications used a literature survey to develop their DQ definition. \num{8} publications adapted a DQ definition to a specific domain. \num{6} publications adapted a DQ definition to a specific dataset type. \num{7} publications used a user study as the seed for the definition. \num{20} publications developed the DQ definition based on a theory.
\end{enumerate}

We review these specialties when discussing research gaps in \autoref{sec:research_gaps}.

\subsection{Threats to Validity}
While the limitations of the underlying literature search are discussed in \autoref{sec:slr_limitations}, this section addresses threats concerning the feature model itself.
The feature model could be refined in several ways: (1) Rather than merely specifying that a definition exists for a specific domain of interest or context of use, the feature model could be refined so that concrete domains or contexts are considered.
We did not make this refinement because the set of possible domains and the set of possible contexts are rather specific to the research done so far.
Rather, we discussed the specific domains and contexts of the publications to illustrate which domains and contexts have already been considered.
(2) Instead of just specifying the contextual relationships of the data, the feature model could classify all the quality dimensions (or aspects) found in quality definitions.
This refinement was not implemented due to the lack of consensus in the literature on how to define quality dimensions.
For example, Batini et al.~\cite{batini2009} define consistency as semantic rules defined over data items, and data must conform to these rules, while Zaveri et al.~\cite{zaveri2015} define consistency as a knowledge base that is free of contradictions.
Another example of disagreement is accessibility.
While Zaveri et al.~\cite{zaveri2015} define accessibility as including ``aspects related to access, authenticity and retrieval of data``, the ISO 25012 standard~\cite{iso2008} defines it as ``the degree to which data can be accessed in a specific context of use, especially by people who need supporting technology or special configuration because of some disability``.
Even without these refinements, the feature model covers all \num{35} DQ definitions identified in our SLR, which shows that it can accommodate their heterogeneity.

We developed this feature model with a \emph{focus on quality definition}.
When the scope is expanded to encompass quality assessment, the definition of quality represents merely the initial stage.
However, it is a crucial stage, as the definition establishes the objective of the entire quality assessment process.
Once the quality goals have been established, the process of quality assessment must begin.
This includes the specification of metrics and quality problems to analyze the current state of quality as well as quality improvement.
Guidelines, such as the FAIR and CARE principles in Wilkinson et al.~\cite{wilkinson2016} and Caroll et al.~\cite{carroll2020}, may be provided in order to supplement these specifications.

A further threat to validity is the classification of the publications in \autoref{tab:slr_publications} according to our feature model. To mitigate this threat, the authors first performed this classification independently and then discussed it to reach a consensus.

\section{Analysis of Research Gaps}
%!TEX root = ../main.tex
\label{sec:research_gaps}
Given the results of our systematic literature review (SLR) and our subsequent classification, we proceed to interpret these results to identify potential research gaps. In our SLR, we identified \num{35} relevant publications that we reviewed in detail. These publications have different aims and benefits, but they all define at least one dimension of DQ. Our classification shows how diverse the landscape of DQ definitions is. On the other hand, there are also various overlaps and strong connections between the publications.
In conclusion, we address (RQ 3), which is stated in \autoref{sec:introduction}:
\begin{enumerate}
    \item[(RQ3)] What research gaps can be identified?
\end{enumerate}

Below we discuss each research gap identified, concluding with an outlined action to address each.

\subsection{Finding Consensus}
\paragraph{Quasi-standards for data quality in general.}
20 of the 35 publications in our SLR define DQ independent of a specific domain of interest (cf.~\autoref{tab:foda_results}), among them the definitions by Wang \& Strong~\cite{wang1996}, Wand \& Wang~\cite{wand1996}, and Batini et al.~\cite{batini2009}, which are additionally independent of dataset type.  
These definitions have been cited extensively (cf.~\autoref{tab:slr_publications}).
Wang \& Strong~\cite{wang1996} is the most frequently cited one in this field.
All these publications provide a comprehensive overview of the quality dimensions defined, with Wand \& Wang~\cite{wand1996} and Batini et al.~\cite{batini2009} offering particularly thorough definitions.
These definitions have become quasi-standards for defining DQ based on dimensions.
Although these publications define DQ based on quality dimensions, they vary considerably in the dimensions selected and in how common dimensions are defined.
This means that there is no agreement in the literature on how to define DQ in general.
In addition, these publications do not cover all contexts of DQ, as quality dimensions related to the social context were not considered in these articles (except for freedom of bias in Wand \& Wang~\cite{wand1996}).
In this respect, these publications appear to be somewhat outdated.
Consequently, there is a need for action to \emph{find consensus on a generally accepted, up-to-date definition of DQ in general}.

\paragraph{Quasi-standards for data quality in certain domains of interest.
}
15 of the 35 publications present a definition of DQ for a specific domain of interest (cf.~\autoref{tab:foda_results}).
Some of these publications are frequently cited and have set a kind of quasi-standard for the particular domain or context
they focus on.
For example, Cai et al.~\cite{cai2015} define the quality of Big Data and
Zaveri et al.~\cite{zaveri2015} define the quality of Linked Data as used on the World Wide Web.
The authors do not only present their definition of DQ but also refer to previous definitions based on the systematic literature review on DQ assessment for Linked Data that they conducted previously.
Kahn et al.~\cite{kahn2016} present a kind of quasi-standard for the quality of electronic health record data.
They reached a community-based consensus on DQ by reviewing DQ publications in their domain, surveying experts and comparing the resulting definition with previously published definitions.
Today, data are generated and analyzed in virtually all domains, and the interest of researchers and practitioners in high quality data is enormous.
Therefore, it is very important to reach a consensus on how to define DQ in a given domain.
Despite of these successful publication on DQ in certain domains, there seems to be no such publications on quasi-standards in other domains such as engineering, political science, and environmental science.
When DQ is defined inconsistently in the literature, then it is difficult for others in the domain to understand and communicate about DQ.
For fields within computer science that work intensively with data, such as ML or more generally artificial intelligence (AI), it is also necessary to understand the DQ requirements, to communicate and harmonize them in the community, and to find a consensus on DQ in this field. A survey of DQ requirements relevant to ML as in~\cite{priestley2023survey} is an important step towards this goal.
\emph{For all domains that work intensively with data, it is necessary  to agree on a domain-specific definition of DQ for a successful quality assessment.}

\paragraph{Standardization of the set of quality dimensions.}
DQ is typically defined as a multifaceted concept that considers a set of quality dimensions. Publications such as Wang \& Strong~\cite{wang1996}, Wand \& Wang~\cite{wand1996}, ISO 25012~\cite{iso2008}, Zaveri et al.~\cite{zaveri2015}, and Ceravolo \& Bellini~\cite{ceravolo2019} present comprehensive catalogs of quality dimensions that broadly capture aspects of DQ.
To better structure the sets of dimensions, the quality dimensions have been classified in different ways.
Over the past decades, many different sets of DQ dimensions have been defined with diverse definitions for each dimension. In Black \& Nederpelt~\cite{black2020}, the authors have compiled a survey of published  DQ dimensions with the ultimate goal to standardize the set of DQ dimensions.
Despite the ongoing standardization efforts, this process remains incomplete. A necessary basis for such a standardization process is a systematic literature review (SLR) on the topic.
A thorough examination of the definitions of quality dimensions in the \num{35} publications that resulted from our SLR reveals that \emph{there is still \underline{no} consensus on which dimensions to include and how to define them}.
For this reason, we have not included individual quality dimensions in our feature model as they lack sufficient standardization at this time.
The standardization process initiated by the DAMA project in~\cite{black2020} should be continued to achieve the ultimate goal \emph{of standardizing the set of DQ dimensions}.

\paragraph{The society as context}
Only 3 of the 35 publications in our SLR consider society as a contextual relationship to DQ (cf.~\autoref{tab:foda_results}): Shanks \& Corbitt~\cite{shanks1999}, Montori et al.~\cite{montori2013}, and Firmani et al.~\cite{firmani2020}.
The 1999 publication by Shanks \& Corbitt~\cite{shanks1999} is an early contribution to the field of DQ that considers social and cultural aspects.
The 2020 publication by Firmani et al.~\cite{firmani2020} focuses on the ethical dimensions such as lack of bias, data provenance, and diversity.
However, none of the 35 publications combines all four contextual relationships, i.e., intrinsic, system, user, and society relationships.
In conclusion, \emph{society-related quality dimensions should be defined in relation to other DQ dimensions.}
In addition, the process of standardizing the set of DQ dimensions should include societal quality dimensions.
This extended definition of DQ may need to be tailored to specific domains.

\paragraph{Adaptation of data quality definitions to new types of data}
There is a growing interest in the quality of rapidly changing, heterogeneous data from different data sources.
Only 6 of the 35 publications explicitly address data change, and 17 address heterogeneous data (cf.~\autoref{tab:foda_results}).
Recent research in the context of Big Data (e.g., Cai et al.~\cite{cai2015}, Klein et al.~\cite{klein2007streaming}, and El Alaoui et al.~\cite{elalaoui2019}) and spatial data in the context of the Internet of Things (e.g., Cai et al.~\cite{cai2015}) clearly demonstrate this.
These publications focus on intrinsic dimensions, such as accuracy and completeness, for these types of data.
Indeed, none of the 6 publications addressing data change considers society as a contextual relationship, i.e., there has been little focus on other contextual relationships such as system, user and society.
It is therefore up to \emph{future research to define a comprehensive range of quality dimensions for these newer forms of data}.
It is also essential to determine which dimensions of the classical definitions of DQ, such as accuracy, completeness, timeliness, understandability, are relevant to these newer forms of data.
It will also be necessary to assess whether additional dimensions are necessary to address the unique characteristics of these newer forms of data.

\paragraph{Reuse of data quality definitions.}
For the definition of DQ in a certain domain or for a specific context of use, authors often recall existing DQ definitions.
According to the provenance of quality definition feature of our classification (cf.~\autoref{tab:foda_results}), 8 of the 35 publications adapted a DQ definition to a specific domain, 6 adapted a DQ definition to a specific dataset type, and only 7 conducted a literature survey as a prerequisite for their own definition.
For example, in their 2019 publication, El Alaoui et al.~\cite{elalaoui2019} reused the quality definition for Big Data in Cai et al.~\cite{cai2015} to define DQ in sentiment analysis.
Chiasera et al.~\cite{chiasera2011} adapted the definition of selected quality dimensions from Batini et al.~\cite{batini2009} to healthcare data.
For defining information quality on the World Wide Web, Kandari et al.~\cite{kandari2011} adopted the quality dimensions that were most frequently mentioned in the literature, especially the definitions from Wang \& Strong~\cite{wang1996}.
Zaveri et al.~\cite{zaveri2015} even conducted a systematic literature survey as a prerequisite for their definition of DQ for Linked Data.
However, it is rather uncommon for existing publications that define DQ dimensions to conduct a systematic literature review.
In many cases, the authors discuss a selection of existing DQ dimensions that are further developed for DQ in general or adapted to a specific domain or context of use.
For example, see Batini et al.~\cite{batini2009} for further development of DQ in general, and Kandari et al.~\cite{kandari2011} for DQ adaptation.
In healthcare, publications are often loosely related to domain-independent quality definitions, e.g., Chiasera et al.~\cite{chiasera2011}, Montori et al.~\cite{montori2013}, and Kahn et al.~\cite{kahn2016}.
Although domain-independent quality definitions are sometimes quoted in the relevant publications, the domain-specific quality definitions can often not be considered as direct refinements of them.
\emph{Specific communities should take a look at the existing (quasi-) standards for DQ in general and in their field of expertise to see to what extent they can be reused.}

\subsection{Comprehensive Definition of Data Quality}

\paragraph{Type of quality definition.}
There are twelve of the 35 publications in our SLR that explicitly define quality dimensions as requirements (cf.~\autoref{tab:foda_results}), such as Wand \& Wang~\cite{wand1996}, Scannapieco et al.~\cite{scannapieco2004}, and Batini et al.~\cite{batini2009} for DQ in general.
The most recent peer-reviewed publications using requirements for quality definition are Zaveri et al.~\cite{zaveri2015} and Firmani et al.~\cite{firmani2020}.
There have also been recent efforts by organizations such as DAMA to provide up-to-date definitions of DQ in terms of requirements (Black \& Nederpelt~\cite{black2020}).
The majority of publications in our SLR define DQ by mentioning quality attributes or dimensions without a declarative definition.
Instead of an explicit definition, these publications use supplements such as quality metrics, examples, or quality problems to explain the quality attributes under consideration.
Given a (standardized) set of DQ dimensions, \emph{each dimension should be defined in a declarative way so that the quality requirements become clear.}

\paragraph{Supplement.}
The definitions of DQ identified in our literature review are typically accompanied by examples, either fictitious or from real-world data sets.
Furthermore, definitions are frequently illustrated with quality problems, metrics, and guidelines, that break down a declarative definition into a list of checks.
Nevertheless, of the 35 publications, only 3 include real-world examples and 9 include fictitious examples, while 23 include no example at all (cf.~\autoref{tab:foda_results}).
The examples that are given are typically to the point and therefore small.
It should be noted that only Strong et al.~\cite{strong1997} report an empirical evaluation of their DQ definition.
The definitions of DQ are otherwise typically not evaluated in an empirical study within the same paper.
In such an empirical study, real-world data sets would be analyzed according to the given quality definition.
\emph{It remains to find publications that evaluate DQ definitions in empirical studies that are based on real-world data.}

\subsection{Guiding Principles for Enhancing Data Quality}
In the context of best practices for data management, two publications have become well-known that present guidelines for enhancing DQ: FAIR and CARE.
The FAIR guiding principles by Wilkinson et al.~\cite{wilkinson2016} aim to improve the transparency, reproducibility, and reusability of research data.
To achieve this goal, data must be findable, accessible, interoperable, and reusable.
The CARE principles by Carroll et al.~\cite{carroll2020} are intended to complement the FAIR principles and refer to collective benefit, authority of control, responsibility, and ethics of data.\\*
Both sets of principles are formulated independently of data representation.
While the FAIR principles are specified for research data, the CARE principles focus on data for governance.
The context of use is generally the reuse of data in an open environment.
The FAIR principles refer to the intrinsic quality of the data, the contextual relationships to the system that provides the data, and to the users of the data.
In contrast, the CARE principles primarily address the relationships with users and society.

Both papers have in common that they are highly cited in the context of DQ, but do not directly define the intended DQ; they rather present guidelines for improving the quality.
Both sets of principles have been developed theoretically and illustrated with real examples.
It is up to future work \emph{to identify a definition of DQ that aligns with the guidelines presented.}

\subsection{Summary of Research Gaps}
In summary, the classification results of the \num{35} relevant publications for defining DQ indicate two types of research gaps:
\begin{enumerate}
    \item \emph{Find consensus:} Actions should be taken to harmonize the set of quality dimensions considered and the definition of each dimension in such a set, both in general and adapted to specific
    application domains.
    \item \emph{Comprehensive definition of DQ:} Actions should be taken to define DQ comprehensively. This definition includes a declarative definition of quality requirements at the core and is supplemented by real-world examples, quality problems that are typically specified by metrics or patterns, and guidelines (e.g., FAIR and CARE) for improving DQ.
\end{enumerate}
Each of these gaps is directly grounded in the quantitative classification results presented in \autoref{tab:foda_results} and \autoref{fig:foda_result}.
Gaps of the first type follow from features that are addressed inconsistently or by only a minority of publications, e.g., only 3 of 35 address society as a contextual relationship, and none combines all four contextual relationships.
Gaps of the second type follow from features that are addressed by only a minority of publications overall, e.g., only 12 of 35 define quality dimensions as requirements, and only 3 include real-world examples.

\section{Conclusion}
%!TEX root = ../main.tex
\label{sec:conclusion}
This paper presents a meta-study in which we screened more than \num{8700} records on data quality (DQ) and identified 35 publications as defining DQ.
We classified these publications to gain a \emph{detailed understanding of how DQ is defined in the literature}, both in general and in specific domains and contexts.
We conducted a systematic literature review to identify which publications contain original, dimension-based definitions of DQ (RQ1). We analyzed the identified set of publications using Feature-Oriented Domain Analysis (FODA) and developed a feature model to answer the question of how to classify the DQ definitions found in this literature (RQ2). Classifying existing DQ definitions using this feature model facilitated the identification of critical research gaps (RQ3).
Our main contributions to answering these research questions are as follows:
\begin{enumerate}
    \item[(RQ1)] As a result of the systematic literature review, we identified 35 relevant publications. Besides the ISO standard on DQ, there are a few highly-cited publications that form quasi-standards for DQ definitions, both for data in general and for specific application domains.
    \item[(RQ2)] We propose a feature model of DQ definitions that covers all 35 identified publications, although their definitions differ widely.
    \item[(RQ3)] In consideration of this classification, we identified two types of research gaps: The considered set of quality dimensions and the definition of each dimension should be harmonized to reach a consensus, both for data in general and adapted to specific application domains. In addition, actions should be taken to define DQ comprehensively, with a declarative definition of quality requirements as the core and supplemented by fictitious and real-world examples, quality problems, and guidelines for improving DQ.
\end{enumerate}

Our findings underscore the value of future consolidation of DQ definitions, both for data in general and for specific application domains and contexts of use.
It has become clear that further research is needed to harmonize the definitions of data quality dimensions in the literature, and in particular to thoroughly address the social and ethical contextual relationships of the quality dimensions.
The resulting data quality definitions should be systematically evaluated in different domains and contexts.
A consistent next step is to systematically investigate existing DQ assurance frameworks that cover not only the definition of DQ but also quality assurance methods and techniques.
In this way, the field can work towards establishing a comprehensive definition of DQ assessment, which will be crucial for the development of high-quality data-intensive systems.

\begin{acks}
We thank Marsha Chechik, Péter Király, Daniel Kurzawe, and Jakob Voß for their valuable feedback and efforts in reviewing our manuscript.
\end{acks}

\bibliographystyle{ACM-Reference-Format}
\bibliography{main}

% \appendix

\end{document}